\documentclass[a4paper, 11pt]{article}
\pdfoutput=1
\usepackage{jcappub} 
\usepackage[T1]{fontenc} 
\usepackage{comment}
\usepackage{bm}

\usepackage{xcolor}
\newcommand{\hl}[1]{\textcolor{magenta}{#1}}
\renewcommand{\hl}[1]{#1}

\newcommand{\Mpl}{M_\mathrm{pl}}
\newcommand{\Lpl}{L_\mathrm{pl}}

\newcommand{\fofR}{f(R)}
\newcommand{\fR}{f_R}
\newcommand{\fRz}{f_{R0}}
\newcommand{\logfRz}{\log_{10}(|\fRz|)}

\newcommand{\Msym}{M_\mathrm{sym}}
\newcommand{\logMsym}{\log_{10}(\Msym/\Mpl)}
\newcommand{\Lc}{L_\mathrm{c}}
\newcommand{\logLc}{\log_{10}(\Lc/\mathrm{kpc})}
\newcommand{\rhoSSB}{\rho_\mathrm{SSB}}

\newcommand{\Mvir}{M_\mathrm{vir}}
\newcommand{\logMvir}{\log_{10}(\Mvir/M_\odot)}
\newcommand{\Rvir}{R_\mathrm{vir}}
\newcommand{\cvir}{c_\mathrm{vir}}
\newcommand{\rnfw}{r_\mathrm{NFW}}
\newcommand{\rhonfw}{\rho_\mathrm{NFW}}
\newcommand{\Mhalo}{M_\mathrm{halo}}

\newcommand{\Mdisc}{M_\mathrm{disc}}
\newcommand{\Rdisc}{R_\mathrm{disc}}
\newcommand{\zdisc}{z_\mathrm{disc}}
\newcommand{\Sigmadisc}{\Sigma_\mathrm{disc}}

\newcommand{\pij}[2]{\varphi_{#1,#2}}
\newcommand{\fij}[2]{f_{#1,#2}}

\title{Accurate Computation of the Screening of Scalar Fifth Forces in Galaxies}

\author[a]{Clare Burrage,}
\author[a]{Bradley March}
\affiliation[a]{School of Physics and Astronomy, University of Nottingham, \\University Park, Nottingham NG7 2RD, UK}
\author[b]{and Aneesh P. Naik}
\affiliation[b]{Institute for Astronomy, University of Edinburgh, \\Royal Observatory, Blackford Hill, Edinburgh EH9 3HJ, UK}
\emailAdd{clare.burrage@nottingham.ac.uk}
\emailAdd{bradley.march@nottingham.ac.uk}
\emailAdd{aneesh.naik@roe.ac.uk}

\abstract{Screening mechanisms allow light scalar fields to dynamically avoid the constraints that come from our lack of observation of a long-range fifth force. Galactic scale tests are of particular interest when the light scalar is introduced to explain the dark matter or dark energy that dominates our cosmology. To date, much of the literature that has studied screening in galaxies has described screening using simplifying approximations. In this work, we calculate numerical solutions for scalar fields with screening mechanisms in galactic contexts, and use these to derive new, precise conditions governing where fifth forces are screened. We show that the commonly used binary screened/unscreened threshold can predict a fifth force signal in situations where a fuller treatment does not, leading us to conclude that existing constraints might be overestimated.
We show that various other approximations of the screening radius provide a more accurate proxy to screening, although they fail to exactly reproduce the true screening surface in certain regions of parameter space. As a demonstration of our scheme, we apply it to an idealised Milky Way and thus identify the region of parameter space in which the solar system is screened.}

\begin{document}
\maketitle
\flushbottom

\section{Introduction}
\label{sec:introduction}

Attempts to resolve long-standing complications within the dark sector often introduce a new scalar field which naturally couples to the gravitational sector, causing modifications to standard gravity, General Relativity (GR), in the form of scalar \textit{fifth forces}. However, such deviations from GR are tightly constrained by lab-based experiments testing the inverse-square law and tests of Solar System dynamics \cite{Adelberger2003}. To avoid such constraints, scalar-tensor theories with \textit{screening} propose that the scalar field is environment-dependent, allowing for modifications to GR on certain scales whilst suppressing fifth forces on other scales \cite{Joyce:2014kja,Koyama:2015vza,Ishak:2018his, BurrageSakstein2018,Brax:2021wcv,CANTATA:2021ktz,Vardanyan:2023jkm}. 

In this paper, we study chameleon \cite{Khoury:2003aq, KhouryWeltman2004} and symmetron \cite{HinterbichlerKhoury2010,Hinterbichler:2011ca} screening models (for related work on other scalar theories see refs.~\cite{Mota:2006fz,Dehnen:1992rr, Gessner:1992flm, Damour:1994zq, Pietroni:2005pv, Olive:2007aj, DGP2000, Vainshtein:1972sx}). Chameleon and symmetron theories introduce a density-dependent scalar potential that screens the fifth force in areas of high density or deep gravitational potential, such as in stellar interiors or the centres of galaxies.

Much of the recent literature has studied the behaviour of these theories on galactic scales, where the screening phenomenon leads to a rich phenomenology, such as equivalence principle violations and other kinematic signatures \cite{Hui+2009, JainVanderPlas2011, Bartlett+2021, Desmond+2018Offsets, Desmond+2018Warps, Desmond+2019, DesmondFerreira2020, Naik+2020Streams, Vikram+2013, Pedersen:2023ina, Naik+2018, Naik+2019, Vikram+2014, LombriserPenarrubia2015, Burrage+2017, Cataldi+2022, O'HareBurrage2018, HogasMortsell2023}. The recent interest in galactic scales is a result of two circumstances: the fact that galaxies inhabit an unconstrained ``desert'' in modified gravity parameter space \cite{Baker+2019}, and the recent advent of vast datasets from various galaxy surveys \hl{\cite{SDSS2023, 2M++2011, Desi2023}}.

However, investigating these scalar fifth forces on galactic scales is a challenging problem: it is difficult to isolate any potential fifth force signal from the numerous astrophysical effects that can obscure it, and forward-modelling the signal involves the solution of a non-linear equation of motion (EoM). As a consequence, much of the literature has relied on approximations to solve the equations of motion, assuming a simplistic setup and dropping terms to simplify the non-linear dynamics. The most common approximation is to simply assume that a galaxy, depending on its mass and environment, is either fully screened or fully unscreened. This disregards the possibility of partial screening, under which the central region of the galaxy is screened while its outskirts remain unscreened. An example of a work using this approximation is that of ref.~\cite{DesmondFerreira2020}, which uses it to derive the current strongest constraints on chameleon $\fofR$ gravity, altogether ruling out the astrophysically-relevant part of parameter space\footnote{In ref.~\cite{DesmondFerreira2020} a Bayesian analysis is used so that a galaxy is assigned a probability of being fully screened, or fully unscreened. It is possible that this captures some of the variation due to galaxies being partially screened, but we are unable to say how much.}. The purpose of the present work is to test these approximations, and the possible impact that relaxing them might have on such constraints.

In this work, we investigate the non-linear nature of screening theories in the galactic regime by utilising a numerical solver, based on refs.~\cite{Puchwein+2013MG-GADGET, Naik+2018, Bose2017(Solver)}, to determine the true scalar field profile inside our static model galaxies. Guided by these field solutions, we obtain empirical rules describing the locations of screening surfaces and compare these results with the results of the common approximations. We discuss the implications of the proper non-linear treatment of these theories and how our findings can impact observables and constraints established using simplified methods. These results will form the foundation for a reevaluation of galaxy-based constraints in the near future, using methods that we have explored and tested in this paper. This reanalysis will enable us to provide a quantitative estimate of the impact of using various standard approximations to screening, compared to our precise numerical methods that consider partial screening.

The outline for this paper is as follows. In section~\ref{sec:theory} we present an overview of the general scalar-tensor theory, as well as our specific chameleon and symmetron models. In section~\ref{sec:numerics} we describe our methods to solve the theories numerically, including the galactic model we use, the mechanics behind the numerical solver, and the criteria used to define the screened and unscreened regions. Section~\ref{sec:galscreening} outlines the results of our investigation. Our conclusions are laid out in section~\ref{sec:conclusion}.


\section{Screened Modified Gravity Theories}
\label{sec:theory}

\subsection{General Scalar-Tensor Model} \label{ssec:General ST theory}

In the present study, our investigation into modified theories of gravity is exclusively focused on scalar-tensor theories. These theories are described by the Einstein frame action, given by
\begin{equation} \label{eq:ST Action}
    S = \frac{c^3}{8\pi G} \int d^4x \sqrt{-g} \left[ 
    \frac{R}{2} 
    - \frac{1}{2} \nabla_\mu\phi\nabla^\mu\phi 
    - V(\phi) \right] 
    + S_m\left[ \tilde{g}_{\mu\nu}, \psi^{SM}_i\right] \,,
\end{equation}
where $g$ is the determinant of the Einstein frame metric, $g_{\mu\nu}$, $R$ is the Ricci scalar, $G$ is the gravitational constant, $\phi$ is a canonically normalised (dimensionless) scalar field with a potential $V(\phi)$, and $S_m$ represents the action for the standard model fields, $\psi_i^{SM}$. The theories also include a conformal coupling to matter, meaning that matter fields move on geodesics of the Jordan frame metric
\begin{equation}
    \tilde{g}_{\mu\nu} = A^2(\phi) g_{\mu\nu} \,.
\end{equation}
Deviations from GR arise from the non-minimal coupling, $A(\phi)$. Although both frames have identical classical observables, for ease of calculation, this analysis focuses solely on the Einstein frame, where there is a direct coupling between matter and the scalar field, and the pure gravitational action does not depend on the scalar field.

Scalar-tensor theories of gravity modify gravitational forces by introducing \emph{fifth forces}, which can be obtained by taking the Newtonian limit of the geodesic equations (see for example ref.~\cite{BurrageSakstein2018}) to arrive at the following expression for the acceleration
\begin{equation} \label{eq:5th force}
    \frac{\bm{a}_5}{c^2}
    = - \beta(\phi)\bm{\nabla}\phi \,,
\end{equation}
where $\beta(\phi) \equiv \mathrm{d}\ln A/\mathrm{d\phi}$. Meanwhile, the scalar field's EoM is derived by extremising the general scalar-tensor action, eq.~\eqref{eq:ST Action}, with an energy-momentum tensor for the matter fields that assumes matter is a non-relativistic, perfect fluid with mass density $\rho$
\begin{equation} \label{eq:general EoM}
    \Box\phi 
    = \frac{dV_{\mathrm{eff}}}{d\phi} 
    \equiv \frac{dV(\phi)}{d\phi} 
    + \frac{8\pi G\rho}{c^2}\beta(\phi) \,,
\end{equation}
where the effective potential, $V_\mathrm{eff}(\phi)$, is given by
\begin{equation} \label{eq:effective potential}
    V_\mathrm{eff}(\phi) 
    \equiv V(\phi) + \frac{8\pi G\rho}{c^2}\ln A(\phi) \,.
\end{equation}

For certain choices of $V(\phi)$ and $A(\phi)$, the effective potential has a stable minimum and this density-dependent EoM admits screening solutions, i.e. solutions in which the fifth force, eq.~\eqref{eq:5th force}, is rendered negligible in certain environments but unleashed elsewhere. Screening can arise via several mechanisms: a small matter coupling, $\beta(\phi)$, a large scalar mass resulting in short-ranged forces, or not all the mass sourcing (or coupling to) the scalar field \cite{BurrageSakstein2018}. 

\subsection{Astrophysical Screening Approximations}
\label{ssec:astro screening approximations}

The crux of this paper is to determine the location of the screened region inside galaxies, defined by what we term the `screening surface': the interface between the screened central region and the unscreened outskirts of a partially screened object. To determine this, we use numerical solutions of scalar EoM. These results will then be compared with the approximate methods that are commonly used in the literature.

One prevalent method to approximate screening surfaces considers a static, spherically-symmetric source, $\delta\rho(r)$, in a constant density background, $\bar{\rho}$, e.g. a star embedded in the galactic medium or a dark matter halo embedded in the cosmological background, as in refs.~\cite{Hui+2009CritPotCalc, Davis+2012CritPotCalc, Sakstein2013CritPotCalc, BurrageSakstein2016CritPotCalc}. For such an object, the screening surface will necessarily be a sphere with radius $r_s$, the `screening radius'. The general EoM, eq.~\eqref{eq:general EoM}, is then solved in a piecewise manner, described in full in appendix~\ref{sec:app chi parameter derivation}, to obtain the expression
\begin{align} \label{eq:chi defn}
    \chi \equiv \frac{\bar{\phi}}{2\beta(\bar{\phi})} &= \frac{4\pi G}{c^2} \int^\infty_{r_s} r\delta\rho(r)dr \,, \\
    \label{eq:critical potential}
     &= -\frac{1}{c^2}\left( \Phi_N(r_s) + r_s \Phi_N'(r_s) \right) \,,
\end{align}
where $\bar\phi = \phi(\bar\rho)$ is the background field solution and $\Phi_N(r)$ is the Newtonian potential. To obtain this $\chi$ parameter, hereby referred to as the \textit{critical potential}, several assumptions are required, specifically that the field reaches its minimum value inside the object, that this minimum value can be neglected, that the theory can be linearised in the region containing the screening radius, that the scalar mass is negligible in this region, and that the coupling function $\beta(\phi)$ is constant throughout. 

If eq.~\eqref{eq:chi defn}, or equally eq.~\eqref{eq:critical potential}, has no solution for $r_s$ this corresponds to a fully unscreened object, i.e. none of the object is screened. Since $\Phi_N<0$ and $\Phi'_N>0$, an object must be fully unscreened when $\chi>|\Phi_N|/c^2$, where the Newtonian potential is measured at the surface of the object. In reverse, this indicates only objects with $\chi<|\Phi_N|/c^2$ are partially screened (or fully screened, i.e. $r_s$ is larger than the extent of the object, in the case $\chi\ll|\Phi_N|/c^2$). 

The critical potential is a starting point for two approaches commonly seen in the literature for approximating the screening of astrophysical objects. In the first approach, the implicit equation~\eqref{eq:chi defn} is solved for the screening radius, which appears as the lower limit of the integral. As we shall see, this can readily be done numerically, but analytic solutions can also be found for certain choices of the density distribution (e.g., for idealised dark matter haloes; as in ref.~\cite{Arnold+2016}).

The second approach, which we refer to as the `binary screening condition', disregards partial screening altogether and classifies objects as either fully screened or fully unscreened by comparing $\chi$ to an estimate of $\Phi_N$ at the surface of the object. Neglecting the contribution of its environment, an object is classed as screened if $|\Phi_N| / c^2 > \chi$, and unscreened otherwise. For example, at the level of current constraints in $\fofR$ gravity, refs.~\cite{Naik+2019, Desmond+2018Offsets, Desmond+2018Warps, DesmondFerreira2020}, objects with deep potentials, e.g. main sequence stars with $|\Phi_N| / c^2 \sim 10^{-6}$ are likely screened, while objects with shallower potentials e.g. dwarf galaxies with $|\Phi_N| / c^2 \sim 10^{-8}$ can still be unscreened. In the context of galaxies, this approach was suggested by ref.~\cite{Cabre+2012}, who proposed using $\Phi_\mathrm{vir}=G\Mvir/\Rvir$, $\Mvir$ and $\Rvir$ will be defined in sec.~\ref{ssec:galaxy model}. The same approach was then subsequently used in a string of recent papers deriving modified gravity constraints on galactic scales (refs.~\cite{Desmond+2018Maps, Desmond+2018Offsets, Desmond+2018Warps, DesmondFerreira2020}). However, it is worth noting that these works also included environmental contributions, classifying objects as screened when $(|\Phi_N| + |\Phi_\mathrm{env.}|) / c^2 > \chi$, where $|\Phi_\mathrm{env.}|$ is the environmental contribution to the local gravitational potential.

One of the main aims of the present work is to compare the results obtained from these two approximate approaches with more accurate results obtained from exact numerical solutions of the scalar equations of motion.

\subsection{Chameleon \texorpdfstring{$\fofR$}{f(R)}}
\label{ssec:Chameleon \fofR}

Chameleon models are a class of scalar-tensor theories that screen primarily via a density-dependent scalar mass, leading to the presence of short-range fifth forces. Screened regions in chameleon models are characterised by flat scalar field profiles, which result in the suppression of the fifth force, eq.~\eqref{eq:5th force}. We choose to work with $\fofR$ chameleon screening because of its simplicity as a model and its ubiquity in the literature.

$\fofR$ theories modify GR by replacing the Ricci scalar, $R$, in the Einstein-Hilbert action with the generalised $R + \fofR$, allowing for higher-order curvature terms in the action (see ref.~\cite{SotiriouFaraoni2010} for a general review),
\begin{equation}
\label{eq:fR action}
    S = \frac{c^3}{16\pi G}\int d^4x \sqrt{-\tilde{g}}\left[R+\fofR\right] 
    + S_m\left[ \tilde{g}_{\mu\nu}, \psi^{SM}_i\right] \,,
\end{equation}
where $R=R(\tilde{g})$. Here, $\fR\equiv\frac{\mathrm{d} \fofR}{\mathrm{d}R}$ acts as the scalar field of the theory. This action can be recast to the general scalar-tensor action eq.~\eqref{eq:ST Action} with the field redefinition \cite{Brax+2008}
\begin{equation}
\label{eq:fR field redefinition}
    \phi=-\frac{\ln(1+\fR)}{2\beta(\phi)} \,, \qquad 
    V(\phi) = \frac{R\fR - \fofR}{2(1+\fR)^2} \qquad
    \mathrm{and} \qquad
    \beta(\phi) = \frac{1}{\sqrt{6}} \,. 
\end{equation}

 For specific choices of the functional form of $\fofR$, the theory can exhibit the chameleon screening mechanism. Several forms of $\fofR$ can achieve this, refs.~\cite{Faulkner+2007f(R)ChameleonModels, Starobinsky2007f(R)ChameleonModels, NavarroVanAcoleyen2007f(R)ChameleonModels}, we adopt the commonly used Hu-Sawicki model, \cite{HuSawicki2007}. In this model
\begin{equation}
\label{eq:fR HS def}
    \fofR = - \frac{am^2}{1 + (R/m^2)^{-b}} \,,
\end{equation}
where $a$ and $b$ are positive dimensionless parameters, and $m$ is a dimensionful parameter. For simplicity, we choose $b=1$, in line with the bulk of the current literature. The two remaining free parameters $a$ and $m$ can be related by requiring that gravity returns to GR + $\Lambda$CDM in the high curvature limit (i.e. $\fofR\approx-2\Lambda$ when $R\gg m^2$), resulting in
\begin{equation}
    am^2 = 2\Lambda \,,
\end{equation}
where $\Lambda$ is the cosmological constant.
This allows us to describe the theory in terms of one free model parameter, $a$ or $m$, which we instead express as the present-day background field value $\fRz$:
\begin{equation}
    a = -\frac{4\Omega_\Lambda^2}{(\Omega_m + 4 \Omega_\Lambda)^2} \frac{1}{\fRz} 
    \quad \mathrm{and} \quad
    m^2 = -\frac{3H_0^2 (\Omega_m + 4\Omega_\Lambda)^2}{2\Omega_\Lambda c^2} \fRz,
\end{equation}
where $\Omega_m$ and $\Omega_\Lambda$ are the cosmological density parameters for matter and dark energy respectively. Throughout this article we assume values of $\Omega_m=0.3$ and $\Omega_\Lambda=0.7$.

Extremising the action eq.~\eqref{eq:fR action} with respect to the metric results in a set of modified Einstein field equations. Taking the trace and then the Newtonian limit, with the assumption $|\fR| \ll 1$ and the quasi-static approximation $|\nabla \fR| \gg \partial_t \fR$, produces the $\fofR$ EoM,
\begin{equation}
\label{eq:fR EoM}
    \nabla^2 \fR = \frac{1}{3}\left(\delta R - \frac{8\pi G}{c^2}\delta\rho\right) \,.
\end{equation}
Here, $\delta\rho$ is the density perturbation and $\delta R$ is the curvature perturbation, given by 
\begin{equation}
\label{eq:fR deltaR}
    \delta R = R_0 \left[\sqrt{\frac{\fRz}{\fR}} - 1\right] \,,
\end{equation}
under the Hu-Sawicki model, with zero subscripts denoting background values today. 

The fifth force acceleration, eq.~\eqref{eq:5th force}, caused by our scalar field is given by
\begin{equation}
\label{eq:fR 5th force}
    \frac{\bm{a}_5}{c^2} = \frac{1}{2}\bm{\nabla} \fR \,,
\end{equation}
where we have assumed $\fR \ll 1$, and the background Compton wavelength is given by 
\begin{equation}
\label{eq: fR background Compton wavelength}
    \lambda_C \approx 32 \sqrt{\frac{\fRz}{10^{-4}}}\, \mathrm{Mpc} \,,
\end{equation}
as in ref.~\cite{Cabre+2012}, based on ref.~\cite{HuSawicki2007}.
Lastly, using our field redefinition eq.~\eqref{eq:fR field redefinition} we can define the critical potential eq.~\eqref{eq:chi defn} in terms of our background scalar field value 
\begin{equation}
\label{eq:fR crit pot}
        \chi = -\frac{3}{2} \fRz  \,.
\end{equation}

\subsection{Symmetron}
\label{ssec:symmetron}

In contrast with chameleon models, where screening arises due to a variable scalar mass, symmetron models screen by suppressing the coupling to matter in dense regions. This is achieved via spontaneous symmetry breaking (SSB), given the correct form of potential and coupling in the scalar-tensor action, eq.~\eqref{eq:ST Action}. In our case, we choose to work with a quartic potential and quadratic coupling,
\begin{align}
    \label{eq:sym potential}
    V(\phi) &= \frac{1}{\Lpl^2}\left[ -\frac{1}{2}\left(\frac{\mu}{\Mpl}\right)^2 \phi^2 + \frac{1}{4}\lambda\phi^4 \right] \,, \\
    \label{eq:sym coupling}
    A(\phi) &= 1 + \frac{1}{2} \left(\frac{\Mpl}{\Msym}\right)^2 \phi^2 \,,
\end{align}  
where $\Mpl^2 \equiv \hbar c / 8\pi G$ is the reduced Planck mass and $\Lpl^2 \equiv 8\pi G \hbar / c^3$ is the reduced Planck length. \hl{In the coupling function above, and throughout the rest of this section, we have assumed that $\phi \ll \Msym/\Mpl$ to suppress higher order terms in $A(\phi)$.} This choice of potential and coupling results in an effective potential eq.~\eqref{eq:effective potential} with the necessary SSB properties. This model has three free parameters: the dimensionless $\lambda$ governing the scalar self-interactions, the mass scale $\Msym$ controlling the strength of the coupling to matter, and $\mu$, representing the bare tachyonic mass of the scalar field.

Extremising the action, eq.~\eqref{eq:ST Action}, with respect to the scalar field gives us the static symmetron EoM, for non-relativistic matter,
\begin{equation}
\label{eq:sym temp EoM}
    \nabla^2\phi = \frac{1}{2L_c^2}\left[ \left( \frac{\rho}{\rhoSSB} - 1 \right)\phi + \lambda\left( \frac{\Mpl}{\mu} \right)^2\phi^3 \right] \,,
\end{equation}
where $L_c$ and $\rhoSSB$ are the background Compton wavelength and SSB density scale, expressed as
\begin{align}
    \label{eq:sym Compton wavelength}
    L_c^2 &= \frac{1}{2}\left(\frac{\Mpl}{\mu}\right)^2 \Lpl^2 \,, \\
    \label{eq:sym rho_SSB}
    \rhoSSB &= \left(\frac{\Msym\mu}{\Mpl^2}\right)^2 \frac{\Mpl}{\Lpl^3} \,.
\end{align}

It is possible to reduce this three-parameter EoM to a two-parameter one by rescaling the field by its vacuum expectation value. We find the vacuum field value by considering solutions to eq.~\eqref{eq:sym temp EoM} in an infinite box of density $\rho$. When $\rho > \rhoSSB$ the field is screened and $\phi=0$. If $\rho < \rhoSSB$, then our scalar field falls into one of two solutions
\begin{equation} \label{eq:sym phi min}
    \phi = \pm \phi_\infty \equiv \pm \phi_\mathrm{vev} \sqrt{1-\frac{\rho}{\rhoSSB}} \,,
\end{equation}
where $\phi_\mathrm{vev}$ is the vacuum expectation value given by
\begin{equation}
    \phi_\mathrm{vev} = \frac{1}{\sqrt{\lambda}} \frac{\mu}{\Mpl} \,.
\end{equation}
Now, rescaling the scalar field, $\varphi=\phi/\phi_\mathrm{vev}$, in eq.~\eqref{eq:sym temp EoM} returns the simpler two-parameter EoM,
\begin{equation}
\label{eq:sym EoM}
    \nabla^2\varphi = \frac{1}{2L_c^2}\left[ \left( \frac{\rho}{\rhoSSB} - 1 \right) + \varphi^2 \right]\varphi \,.
\end{equation}

The acceleration a test particle experiences due to the scalar fifth force is given by
\begin{equation}
\label{eq:sym 5th force}
    \frac{\bm{a}_5}{c^2} = - \phi_\mathrm{vev}^2 \left(\frac{\Mpl}{\Msym}\right)^2 \varphi\bm{\nabla}\varphi \,.
\end{equation}
Note that despite reducing the EoM to two parameters, the theory remains a three-parameter one as the scaling of the fifth force with $\phi_\mathrm{vev}^2$ reintroduces a dependence on $\lambda$. The critical potential eq.~\eqref{eq:critical potential} for this symmetron model is 
\begin{equation}
\label{eq:sym crit pot}
    \chi = \frac{1}{2} \left(\frac{\Msym}{\Mpl}\right)^2 \,.
\end{equation}
Note that $\chi$ depends only on $\Msym$. However, as we shall see later, screening is a function of both $\Msym$ and $\Lc$. This discrepancy is a result of the approximations used to derive the general form of $\chi$, eq.~\eqref{eq:chi defn}\hl{; specifically a combination of assuming the field can be linearised at the screening surface, the background scalar mass is light and thus negligible, and that $\beta(\phi)$ is constant throughout.}

The symmetry-breaking nature of the symmetron provides a third possible approximation for the screening surface --- in addition to the two described in section \ref{ssec:astro screening approximations} for scalar-tensor theories in general. One can assume that the screening surface universally corresponds to the surface where $\rho(r=r_s)=\rhoSSB$, neglecting other relevant factors such as the coupling strength and Compton wavelength of the theory. We will also examine the efficacy of this approximation in section \ref{sec:galscreening}.


\section{Numerical Screening Solutions} \label{sec:numerics}

This section describes the numerical methods we use to obtain isolated galaxy solutions for the scalar field equations of motion under the chameleon $\fofR$, eq.~\eqref{eq:fR EoM}, and symmetron theories, eq.~\eqref{eq:sym EoM}. Section~\ref{ssec:galaxy model} describes our choice of galaxy model, section~\ref{ssec:solver} then describes our numerical solver for the equations of motion, and section~\ref{ssec:Screening conditions} outlines the conditions we use to identify the screening surface within our scalar field solutions. 

\begin{figure}[tbp]
\centering 
\includegraphics[width=\textwidth, trim=0 0 0 0, clip]{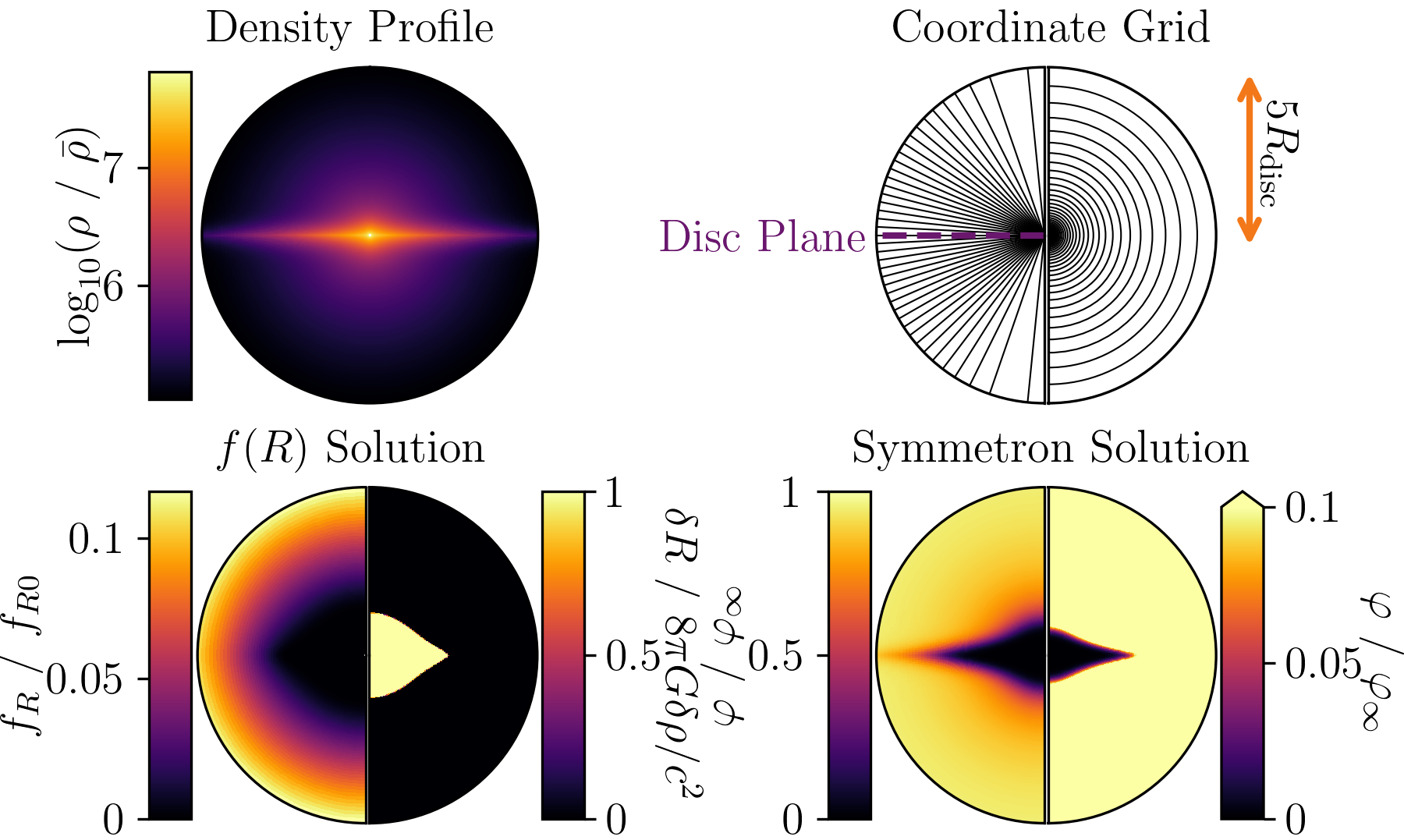}
\hfill
\caption{\label{fig:1. density and field solns} Various panels illustrating the different stages of our numerical procedure, as described throughout section~\ref{sec:numerics}. All plots are zoomed in to $5\Rdisc$, showing only the disc-dominated central region. \textbf{Top left}: input density profile, for an axisymmetric galaxy with $\Mvir=10^{11.5}\,M_\odot$, rescaled by the cosmic mean density. \textbf{Top right}: coordinate grid, showing every \textit{fifth} radial and azimuthal angle gridline. \textbf{Bottom left, left half}: $\fR$ field profile rescaled by the background field value, $|\fRz| = 10^{-6.4}$. \textbf{Bottom left, right half}: the curvature-density ratio, eq.~\eqref{eq:fR screening condition}. \textbf{Bottom right, left half}: symmetron field profile rescaled by its background field value, for the input parameters $\Msym = 10^{-4.5}\,\Mpl$ and $\Lc = 10^{-1}\,\mathrm{kpc}$. \textbf{Bottom right, right half}: truncated scalar field profile to illustrate the symmetron screening condition, eq.~\eqref{eq:sym screening condition}.}
\end{figure}

Our methods are summarised in figure~\ref{fig:1. density and field solns}, the individual panels of which show a typical density profile under our galaxy model, the coordinate grid of our scalar field solver, the scalar field solutions for the same galaxy, and the screening surfaces identified from the scalar field solutions. These various panels of the figure will be examined in more detail in the corresponding parts of this section.

\subsection{Galactic Model} \label{ssec:galaxy model}

Given the $\fofR$ and symmetron EoMs, equations~\eqref{eq:fR EoM} and \eqref{eq:sym EoM} respectively, we can solve the scalar field profile given a suitable galactic density model. For this, we use a two-component axisymmetric model comprising a dark matter halo and a galactic disc (note that we are limiting our study to late-type \hl{spiral disc galaxies, not elliptical or dwarf spheroidal galaxies}). We model the dark matter halo with a Navarro-Frenk-White (NFW) profile \cite{NFW1996}
\begin{equation}
\label{eq:NFW profile}
    \rho(r) = \frac{\rhonfw}{\dfrac{r}{\rnfw} \left(1+\dfrac{r}{\rnfw}\right)^2} \,,
\end{equation}
where $r$ is the spherical radial coordinate, and $\rhonfw$ and $\rnfw$ are the characteristic density and length scale. The galactic disc is represented by a simple double exponential profile\footnote{Using a continuous density profile for a discrete distribution of stars is technically only correct if the Compton wavelength of our theory is larger than the typical separation of stellar objects. This is true for all $\fofR$ parameters we consider, but in the symmetron case our adopted parameter range approaches the parsec scale at its extreme end.} \cite{BinneyTremaine2008}
\begin{equation}
\label{eq:disc profile}
    \rho(R,z) = \frac{\Sigmadisc}{2\zdisc} e^{-R/\Rdisc}e^{-|z|/\zdisc} \,,
\end{equation}
where $R$ represents the cylindrical radial coordinate and $z$ represents the 
coordinate perpendicular to the galactic disc, and $\Sigmadisc$, $\Rdisc$ and $\zdisc$ are the characteristic density and length scales. In actuality, a more sophisticated galaxy model (e.g. \cite{McMillan2017}) could also include various other components, such as central bulges, stellar haloes, and multiple discs; however, we have chosen to exclude these for the sake of keeping our model as simple as possible while still capturing the key characteristics. 

Including these density components will introduce five extra input parameters used to scale the density profiles: $\rhonfw$, $\rnfw$, $\Sigmadisc$, $\Rdisc$ and $\zdisc$. Nonetheless, for the remainder of this article, we shall quote our galactic model in terms of one input parameter: the virial mass, $\Mvir$. This is defined as the mass enclosed within the virial radius, $\Rvir$, the radius within which the mean density inside is 200 times the cosmological critical density. We reduce the five-dimensional parameter space into a one-dimensional space by using empirical relations between the various parameters. In doing so, we typically adopt median values on the relations, ignoring any scatter. Thus, for a given $\Mvir$, the galaxy we model is in some sense a maximally typical one. We describe this process in further detail in appendix~\ref{sec:App. galaxy pipeline}. The top left panel of figure~\ref{fig:1. density and field solns} shows an example density profile for a galaxy with $\Mvir=10^{11.5}\,M_\odot$, corresponding to the following values for the parameters appearing in our density model: $\rhonfw=5.2\times10^{6}\,M_\odot\,\mathrm{kpc}^{-3}$, $\rnfw=15\,\mathrm{kpc}$, $\Sigmadisc=6.4\times10^{8}\,M_\odot\,\mathrm{kpc}^{-2}$, $\Rdisc=1.6\,\mathrm{kpc}$ and $\zdisc=0.26\,\mathrm{kpc}$.

The adoption of the NFW profile for the dark matter halo presents two numerical challenges: the central density singularity and the logarithmically divergent mass. To resolve these issues, we set a constant density within the central 50\,pc, and introduce an outer cutoff at 2.2$\Rvir$. A sharp cutoff at the halo's edge is well motivated astrophysically, e.g. ref.~\cite{Deason2020}. \hl{We tested our analysis for various inner and outer cutoffs. In changing the inner cutoff neither the derived position of the screening surfaces, nor the conclusions we make are altered. We found that increasing the outer cutoff moved the screening surface farther outward, primarily due to the increase in the total mass of the system. However, we find this shift is mirrored in true and approximate solutions, and thus our conclusions remain robust.}

In this work, we assume that the galaxy is embedded in a constant-density background equal to the cosmological critical density. Therefore, we do not consider environmental effects, such as inhomogeneities in the matter distribution throughout the Universe, which can manifest as regions of lower density (voids) and higher density (clusters, walls, and filaments), as well as neighbouring galaxies and satellites. This simplification enables us to better investigate other variables that impact screening in galaxies. Overall, higher-density environments tend to enhance screening effects, resulting in more fully screened galaxies and larger screened regions in partially screened galaxies, while the opposite is true for lower-density environments. We comment on where this will impact our results throughout section~\ref{sec:galscreening}, and plan to account for these effects in follow-up work. 

\subsection{Numerical Solver} \label{ssec:solver}

Using the axisymmetric galaxy density profiles from the previous section, we solve the $\fofR$ and symmetron equations of motion, given by eqs.~\eqref{eq:fR EoM} and \eqref{eq:sym EoM} respectively. We do this using a numerical solver based on that employed by ref.~\cite{Naik+2018}, updated to incorporate some optimisations suggested by \cite{Bose2017(Solver)} and to include the symmetron. The solver is ultimately a two-dimensional version of the $\fofR$ solver used in the cosmological simulation code \textsc{mg-gadget} \cite{Puchwein+2013MG-GADGET}: it constructs a coordinate grid, discretises the EoM on this grid, then solves this discretised EoM using an iterative Gauss-Seidel method. 

Although a cylindrical polar coordinate system might seem the natural choice for an axisymmetric system, we use a two-dimensional spherical polar system (neglecting the azimuthal coordinate). This is because the radial range of our grid should stretch from the galactic centre to several Compton wavelengths (eqs.~\eqref{eq: fR background Compton wavelength} and \eqref{eq:sym Compton wavelength}) beyond the galaxy's edge; the dynamic range is such that a spherical coordinate system better enables us to use a variable grid resolution with smaller cells in the central region and larger cells further out. Specifically, we use a logarithmic radial discretisation, i.e. cells are equally spaced in the coordinate \hl{$u \equiv \ln (r/\mathrm{[chosen\ length\ scale]})$}, producing higher resolution at smaller radii, as required. 
Similarly, in the polar direction, $\theta$, because the galaxy has a disc, we need a higher resolution near the disc-plane $\theta \approx \pi / 2$ than at the poles. We can achieve this by discretising in the coordinate $v \equiv \cos \theta$. In principle, one could assume mirror symmetry about the disc plane and restrict $v$ to the range $[0, 1]$, but we are often interested in the behaviour of the scalar field in the disc plane, and so wish to avoid pushing the disc plane to the boundary of the grid. We thus discretise $v$ in the range $[-1, 1]$. A visualisation of the coordinate grid, out to a maximum radius of $5\Rdisc$, is given in the top-right panel of figure~\ref{fig:1. density and field solns}.

In $u-v$ coordinates, the Laplace operator acting on a function $f$ (dropping the azimuthal term) is
\begin{equation}
\label{eq:LaplaceXY}
    D \equiv \nabla^2 f = \frac{1}{r^3}\frac{\partial}{\partial u}\left(r \frac{\partial f}{\partial u}\right) + \frac{1}{r^2}\frac{\partial}{\partial v}\left(\hl{(1-v^2)} \frac{\partial f}{\partial v}\right) \,.
\end{equation}
For later convenience, we have only expressed the derivatives in terms of \hl{$u$, while keeping other factors that contain $r$, as the conversion to $u$ is straightforward.}

We construct a regular grid in $u-v$ space. We use 512 radial ($u$) cells in the range $[0.05, 10^4]$\,kpc, and 201 angular ($v$) cells. The maximum radial extent of $10^4$\,kpc is several times larger than any \hl{galaxy scale considered in this work, so the field can reach the background value outside of the galaxy.} By using an odd number of angular cells, we ensure that one row of the grid is centred on the disc plane. Each grid cell is labelled with indices $(i, j)$, where $i$ represents the $u$-coordinate and $j$ represents the $v$-coordinate, i.e. $u=u_i$ and $v=v_j$ at cell $(i, j)$. All other quantities (density, scalar field etc.) can be discretised on this grid. For example, we can write $\rho_{ij}$ to represent the density at the \textit{centre} of cell $(i, j)$.

The Laplace operator acting on a function $f$, eq.~\eqref{eq:LaplaceXY}, can then be discretised as
\begin{equation}
\label{eq:Dij}
\begin{split}
    D_{ij} & = \frac{1}{h_u^2 r_i^3}\left[\fij{i+1}{j}r_{i+1/2} + \fij{i-1}{j}r_{i-1/2} - f_{ij}\left(r_{i+1/2} + r_{i-1/2}\right)\right] \\ 
    & + \frac{1}{h_v^2 r_i^2}\left[\fij{i}{j+1}s_{j+1/2}^2 + \fij{i}{j-1}s_{j-1/2}^2 - f_{ij}\left(s_{j+1/2}^2 + s_{j-1/2}^2\right)\right] \,,
\end{split}
\end{equation}
where $h_u$ and $h_v$ are the (constant) grid spacings in $u$ and $v$ respectively, and $s \equiv \sin\theta$, so that $s_j^2 = 1 - v_j^2$. Note that half-integer subscripts (e.g. $r_{i+1/2}$) indicate quantities evaluated at the cell edges rather than the cell centres.

We can now proceed to discretising the EoMs, eqs.~\eqref{eq:fR EoM} and \eqref{eq:sym EoM}. In the $\fofR$ case, we follow reference~\cite{Bose2017(Solver)} in redefining the scalar field as $\varphi \equiv (\fR / \fRz)^{1/2}$. Consequently, the discretised forms of the symmetron and $\fofR$ EoM both turn out to be cubic equations of the form
\begin{equation}
\label{eq:EoMDiscrete}
    \mathcal{L}_{ij} \equiv \varphi_{ij}^3 + p_{ij}\varphi_{ij} + q_{ij} = 0 \,,
\end{equation}
where the coefficients $p_{ij}$ and $q_{ij}$ are given in the $\fofR$ case by
\begin{equation}
\begin{split}
    p_{ij} & \equiv -\frac{1}{A_{ij}}\left(B_{ij}(\varphi^2) + \frac{8\pi G \delta\rho_{ij}}{3c^2 \fRz} + \frac{R_0}{3\fRz}\right) \,; \\
    q_{ij} & \equiv \frac{1}{A_{ij}}\frac{R_0}{3\fRz} \,,
\end{split}
\end{equation}
and in the symmetron case by
\begin{equation}
\begin{split}
    p_{ij} & \equiv \frac{\rho_{ij}}{\rhoSSB} + 2 L_c^2 A_{ij} - 1 \,; \\
    q_{ij} & \equiv - 2L_c^2 B_{ij}(\varphi) \,,
\end{split}
\end{equation}
and the objects $A_{ij}$ and $B_{ij}$ are given in turn by
\begin{equation}
\begin{split}
    A_{ij} & = \frac{r_{i+1/2} + r_{i-1/2}}{r_i^3 h_u^2} + \frac{s_{j+1/2}^2 + s_{j-1/2}^2}{r_i^2 h_v^2} \,; \\
    B_{ij}(\psi) & = \frac{r_{i+1/2}\psi_{i+1, j} + r_{i-1/2}\psi_{i-1, j}}{r_i^3 h_u^2} + \frac{s_{j+1/2}^2 \psi_{i, j+1} + s_{j-1/2}^2 \psi_{i, j-1}}{r_i^2 h_v^2} \,, 
\end{split}
\end{equation}
where $\psi$ represents the argument of $B_{i,j}$, i.e. $\varphi^2$ in the $\fofR$ case and $\varphi$ in the symmetron case.

Before proceeding to solve eq.~\eqref{eq:EoMDiscrete} we must specify our boundary conditions and initial field profile guesses, which we will iteratively converge to the true solution. Our grid has four boundaries: $v=-1, 1$ (i.e. $\theta=\pi,0$) and $u = \ln r_\mathrm{min}, \ln r_\mathrm{max}$. At the first three, we impose $\partial \varphi / \partial v, \partial \varphi / \partial u = 0$. At the outer radial boundary, we instead impose $\varphi = \varphi_\infty$, i.e. the cosmic background value of the scalar field. In the $\fofR$ case, this is simply $\fRz$; in the symmetron case, this is the field value that minimises the effective potential eq.~\eqref{eq:sym phi min} evaluated at the cosmic matter density. For initial guesses, in the $\fofR$ case we simply set $\varphi=1$ everywhere, and allow the solver to converge on the correct profile. For the symmetron case, we speed up the convergence by setting the field to 1 everywhere, except in the region where $\rho>\rhoSSB$ which is assumed to be screened and thus set to 0. 

We solve eq.~\eqref{eq:EoMDiscrete} using a Gauss-Seidel iteration method. Here, our 2D grid is decomposed into a chequerboard pattern, i.e. we alternately assign cells with `red' and `black' labels. Each iteration $n$ is then comprised of two halves: a red sweep and a black sweep. During the `red sweep', the scalar field value is changed in all of the red cells using a Newton-Raphson update,
\begin{equation}
    \pij{i}{j}^n = \pij{i}{j}^{n-1} - \frac{\mathcal{L}_{i, j}}{\mathcal{L}'_{i, j}} \,,
\end{equation}
where $\mathcal{L}' \equiv \partial \mathcal{L} / \partial \varphi  = 3 \varphi^2 + p$. Note that $\mathcal{L}_{i, j}$ here is evaluated using $\varphi^{n-1}$, i.e. the values of $\varphi$ calculated at the previous Gauss-Seidel iteration (or the initial guess in the case $n=1$). 

Next, we perform a `black sweep' in which the scalar field is updated in all of the black cells using the same expression. There is however one key difference in the black sweep: rather than using the leftover values from the previous Gauss-Seidel iteration ($n-1$) as was done during the red sweep, one uses the neighbouring red cell values that have just been updated during the red sweep, i.e. in the first half of iteration $n$.

This iterative process repeats, updating the field values at each grid point until a converged solution is obtained. The iteration stops when the difference between the field in the previous step and the current one is less than a given threshold. In the $\fofR$ case we use a threshold of $10^{-7}$, and in the symmetron case we use $10^{-14}$. We find these values give robust, converged results. 

The left semicircles in the bottom row of figure~\ref{fig:1. density and field solns} show outputs of the solver for the same example galaxy shown in the upper left panel of the figure, with a mass $\Mvir=10^{11.5}\,M_\odot$, theory parameters of $\fRz=-10^{-6.4}$ in the $\fofR$ case, $\Msym=10^{-4.5}\,\Mpl$ and $\Lc=10^{-1}\,\mathrm{kpc}$ in the symmetron case. In both cases, the field solutions are plotted to a maximum radius of $5\Rdisc$.

\subsection{Screening Conditions}\label{ssec:Screening conditions}

This section aims to establish a universal and reliable screening condition to locate the (generally non-spherical) screening surface, $r_s(\theta)$, in a galaxy's scalar field profile. We will find that our conditions to calculate $r_s(\theta)$ are given by 
\begin{subequations} \label{eq:screening conditions}
\begin{align}
    \left.\frac{c^2\delta R(r,\theta)}{8\pi G\delta\rho(r,\theta)}\right|_{r=r_s(\theta)} = 0.9 \,, \label{eq:fR screening condition} \\
    \left.\frac{\varphi(r,\theta)}{\varphi_\infty}\right|_{r=r_s(\theta)} = 0.1 \,, \label{eq:sym screening condition} 
\end{align}
\end{subequations}
for the $\fofR$ and symmetron models respectively.
In other words, for a given angle, the screening surface is the surface at which the $\fofR$ \textit{curvature-density} ratio or the symmetron field value matches a given threshold. The right sides of the bottom row panels in figure~\ref{fig:1. density and field solns} display the $\fofR$ curvature-density ratio and the symmetron field with a colour map truncated at the threshold. The sharp transition in both of these cases vindicates our chosen threshold. \hl{We arrived at the screening conditions presented in this section, after considering a variety of different properties of the scalar field solutions, for example the fifth force profiles. The screening conditions discussed here are the ones that could be most consistently applied across the range of model and galaxy parameters explored in this work.}

In the $\fofR$ case, the screening condition, eq.~\eqref{eq:fR screening condition}, is a threshold on the curvature-density ratio, motivated by the two terms on the right-hand side of the $\fofR$ EoM eq.~\eqref{eq:fR EoM}. Specifically, inside the screened region the field is flat and the Laplacian is zero, thus the two terms on the right-hand side of the EoM are approximately equal in magnitude, and the curvature-density ratio is unity. Conversely, in the unscreened regime, the field adopts its background value, causing $\delta R$, eq.~\eqref{eq:fR deltaR}, and thus the curvature-density ratio to become zero.
We find that the curvature-density ratio provides a clearer demarcation between the screened and unscreened region than the scalar field, as can be seen in the bottom left plot of figure~\ref{fig:1. density and field solns}. Nevertheless, there are still instances where the transition is more gradual, particularly for smaller values of $|\fRz|$. We find that a threshold on the curvature-density of 0.9 correctly separates the screened/unscreened regions and produces universal behaviour across the whole parameter space.

For the symmetron condition, we use a threshold value on the field itself. This is motivated by the coupling function, eq.~\eqref{eq:sym coupling}, having a first derivative proportional to the scalar field; therefore the density-dependent term in the EoM is then also proportional to the scalar field. Setting this threshold as a ratio of the background scalar field allows for a screening condition that can be universally applied over the whole parameter space. We determine that a threshold value of 10\% of the background scalar field value provides consistent results, with the scalar field profile being visually flat, and thus suppressing fifth forces, within a substantial region of $r_s$.

\begin{figure}[tbp]
\centering 
\includegraphics[width=\textwidth, trim=0 0 0 0, clip]{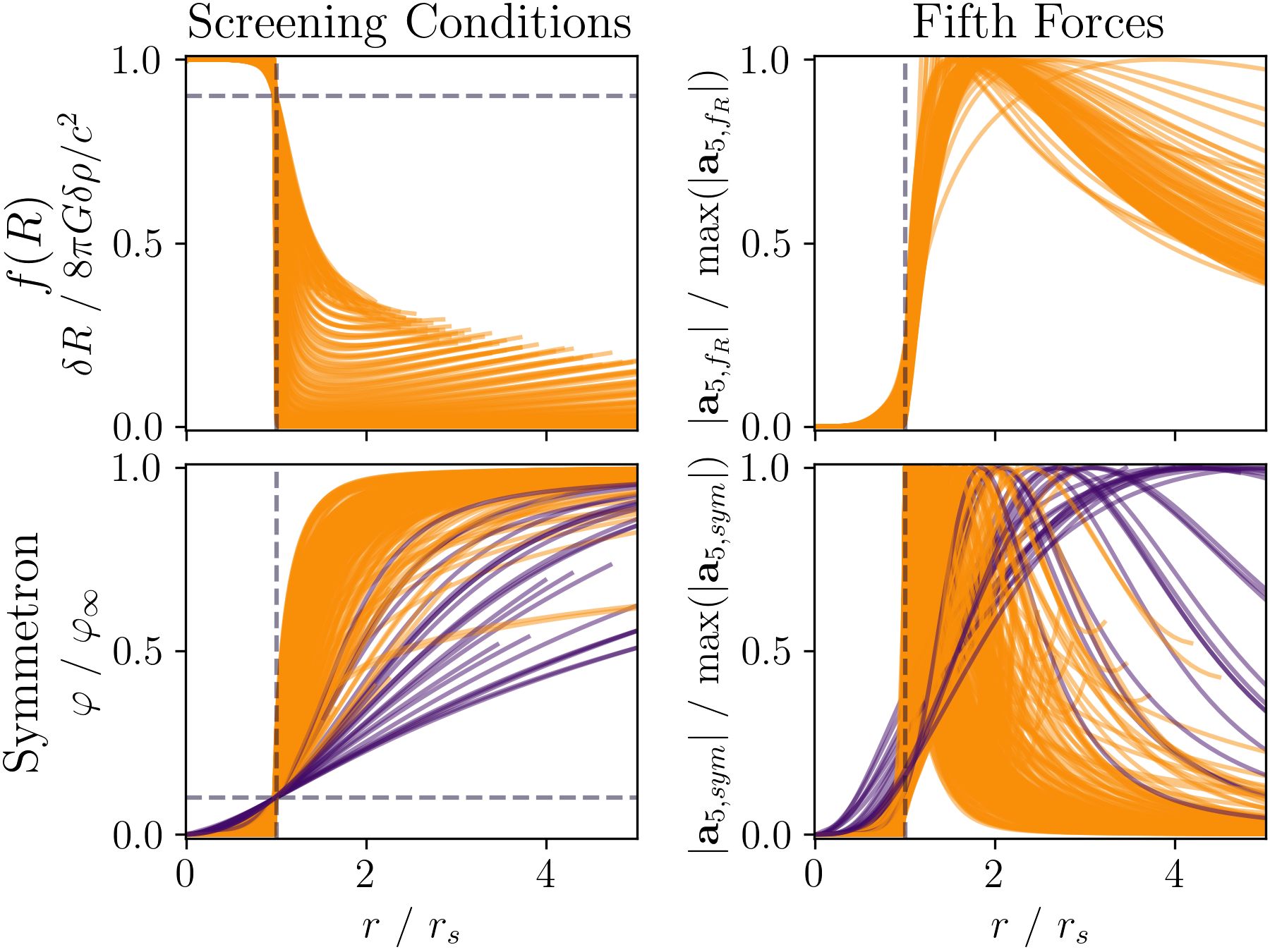}
\hfill
\caption{\label{fig:2 stacked screening conditions} 
Values of the $\fofR$ curvature-density ratio (\textbf{\hl{top left}}), the $\fofR$ fifth force (\textbf{\hl{top right}}, the symmetron field (\textbf{\hl{bottom left}}) and the symmetron fifth force (\textbf{\hl{bottom right})}, in the plane of the galactic disc, plotted against radial coordinates rescaled by $r_s$ (the radius at which the screening surface meets the disc plane), across a range of theory parameters and galaxy masses. In the $\fofR$ case, $\logfRz \in [-8, -5]$ and $\logMvir \in [10, 13.5]$, both in increments of 0.2. In the symmetron case, $\logMsym \in [-6.5,-3]$, $\logLc \in [-3, 3]$ and $\logMvir \in [10, 13.5]$, all in increments of 0.5.  Profiles are cut off at the extent of the galaxy, $2.2\Rvir$. Dashed vertical lines represent the location of the screening radius. Dashed horizontal lines correspond to the screening condition threshold. Purple lines on the symmetron plot indicate solutions that evade the fully unscreened central scalar field definition but are properly considered fully unscreened by the central Laplacian definition.}
\end{figure}

Figure~\ref{fig:2 stacked screening conditions} shows the value of the $\fofR$ curvature-density parameter and the symmetron field along the plane of the galactic disc, as well as the magnitude of the fifth forces of both theories. We rescale the radial coordinates by the derived $r_s$ along the disc plane to better visualise the behaviour over the whole parameter space. Taking figures~\ref{fig:1. density and field solns} and \ref{fig:2 stacked screening conditions} together, we are able to justify our choice of thresholds, namely 0.9 for the curvature-density parameter and 0.1 for the symmetron field, finding that this choice allows for consistent results, that suppress the fifth forces inside of the screened region, across our full range of parameters.

We classify a galaxy as fully screened if the screening surface determined according to eqs.~\eqref{eq:screening conditions} is greater than the extent of the galaxy, or if SSB has not yet occurred in the symmetron case, i.e. $\bar{\rho} > \rhoSSB$. Conversely, a galaxy is considered fully unscreened if there is no screened region within the galaxy, i.e., if the conditions given by eq.~\eqref{eq:screening conditions} are not satisfied anywhere. However, we also find a small subset of galaxies for which the screening conditions are satisfied somewhere but nonetheless have non-negligible fifth forces in their central regions. To catch these cases, we consider the innermost field value. In the $\fofR$ case, solutions satisfying $\fR(r\approx 0) > 10^{-3}\fRz$ are fully unscreened. We find that an analogous threshold on the field central field value can not be used in the symmetron case, as some field solutions yield small central field values but quickly grow away from the centre, giving an unscreened central region. So, for the symmetron, we instead use the Laplacian of the field: solutions satisfying the condition $\nabla^2\varphi(r\approx 0) > 10^{-1} \mathrm{max}\left(\nabla^2\varphi\right)$ are fully unscreened. 

The symmetron solutions that avoid the central field threshold condition (i.e. satisfying $\phi(r\approx 0) < 10^{-3}\phi_\infty$) but are correctly classified as fully unscreened by the central Laplacian threshold condition are highlighted in figure~\ref{fig:2 stacked screening conditions}. Here we see that these edge case field profiles grow to be large well within the screening radius, causing a non-zero fifth force in this region, as shown in the bottom panel.


\section{The Screening of Galaxies}
\label{sec:galscreening}

\subsection{Position and Shape of Screening Surfaces}
\label{ssec: partial screening}

\begin{figure}[tbp]
\centering 
\includegraphics[width=\textwidth, trim=0 0 0 0, clip]{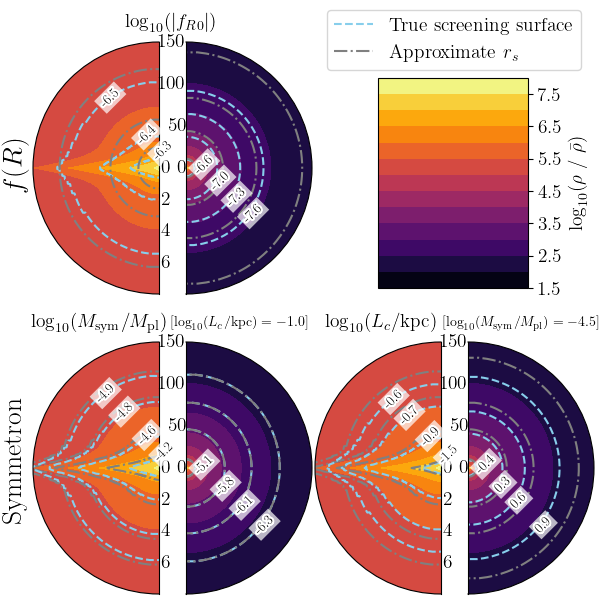}
\hfill
\caption{\label{fig:3. Partial Screening} Screening surfaces across a range of theory parameters, in a galaxy with mass $\Mvir=10^{11.5}\,M_\odot$, for $\fofR$ (\textbf{top}) and symmetron (\textbf{bottom}) models. In the symmetron case, we vary $\Msym$ holding $\Lc$ fixed (\textbf{bottom left}) and vary $\Lc$ holding $\Msym$ fixed (\textbf{bottom right}). Sky blue dashed lines are the calculated screening surfaces with theory parameters labelled. Grey dashed-dotted lines are approximate solutions: numerically integrating the critical potential \eqref{eq:chi defn} in the $\fofR$ case; isodensity contours equal to $\rhoSSB$ in the symmetron case. In each case, the right-hand semi-circle plots the radial range of the galaxy out to \hl{$10\rnfw$ (approximately 150\,kpc)}, while the left-hand semi-circle zooms into the central region, out to $5\Rdisc$ (approximately 8\,kpc). The coloured background regions represent the adopted matter density profile, plotted as a fraction of the cosmic mean density.}
\end{figure}

Figure~\ref{fig:3. Partial Screening} shows an array of solutions for the screening surface (dashed lines), over a range of $\fofR$ and symmetron parameters, for a galaxy with $\Mvir=10^{11.5}\,M_\odot$. Generally, decreasing the adopted value of $|\fRz|$ or $\Msym$, or increasing $\Lc$, pushes the screening surface further out. We separate these solutions into those for which the screening surfaces lie within the disc-dominated region (left-hand semi-circles) and the halo-dominated region (right-hand semi-circles). The figure shows that screening surfaces that occur in the disc-dominated region are elongated along the galactic disc, whereas larger screening surfaces are spherical and trace the dark matter component. 

We also include approximate screening surfaces (dashed-dotted lines) for comparison. In the case of $\fofR$, the approximate screening surface is computed by numerically solving eq.~\eqref{eq:chi defn}, where $\chi$ is given by the $\fofR$ critical potential, eq.~\eqref{eq:fR crit pot}, and $\delta\rho(r)$ is the galactic density averaged over spherical shells. The approximate symmetron screening surfaces are isodensity contours with a value equal to the symmetry breaking scale, i.e. $\rho(r=r_s(\theta))=\rhoSSB$. 

In the $\fofR$ case (top left panel), the true screening surfaces (dashed lines) in the disc-dominated region follow the isodensity contours away from the disc plane, but deviate nearer the disc, where the isodensity contours are flattened to a larger extent. We find that the screening surfaces lie somewhere between the highly flattened isodensity contours of our galactic model and the isopotential contours (not shown), which are visually more spherical than the screening surfaces.

The non-spherical nature of the screening surface in the disc-dominated region of the galaxy (left-hand side of the $\fofR$ plot) indicates that using approximations (dashed-dotted lines) that result in spherical screening radii, such as eq.~\eqref{eq:chi defn}, can fail to capture the true nature of the screened region. This is especially relevant when searching for observables along the plane of the disc, where $r_s$ will be underestimated as the spherical screening surface meets the disc plane at a smaller radius than the true, flattened screening surface. As a result, for a given value of the theory parameter, one may expect the presence of fifth force observables at a certain location in the disc plane because it lies outside the assumed spherical screening surface, whereas, in reality, the location lies within the true, flattened, screening surface, and so the fifth force is suppressed. This in turn could lead to exaggerated constraints on the theory parameters.

In the halo-dominated region of the galaxy (right-hand side of the $\fofR$ plot) we see the approximation overestimates the screening radius for smaller values of $|\fRz|$ --- corresponding to larger values of $r_s$\hl{, and smaller Compton wavelengths}. This leads to an overestimation of the amount of dark matter that is screened by this approximation. This behaviour is likely due to a breakdown of the approximations utilised to derive the critical potential at large $r_s$ values, namely that the field can be linearised around the screening surface, and that the scalar mass is negligible in this region. \hl{Specifically, when the Compton wavelength is smaller than the extent of the galaxy, not all the mass within the galaxy impacts the field profile, whereas the approximation, which assumes negligible scalar mass and thus infinite Compton wavelength, responds to mass in the extremities of the galaxy.} The effect that this overestimation of $r_s$ has on observables is discussed further in section~\ref{subsec:massscreened}.

By contrast with the $\fofR$ screening surfaces, the screening surfaces under the symmetron theory (bottom plots) follow the isodensity contours very closely. However, this behaviour breaks down in the large $\Lc$, large $\Msym$ regime, as seen with the larger screening radii in the left-sided symmetron plots, where the screening surface fails to trace the finer details of the isodensity contour, especially near the disc plane. The reasoning for this breakdown is due to the physics associated with each parameter. The Compton wavelength, $\Lc$, is the characteristic scale on which the field can vary, i.e. the field will not respond to changes on scales much smaller than the Compton wavelength, but in the large $\Lc$ regime the scale height of the disc profile is much smaller than $\Lc$. In other words, a larger Compton wavelength responds to mass averaged over a larger distance, causing the field profile to smooth out the finer details of the disc density profile. Meanwhile, for the coupling constant, $\Msym$, the coupling function, and thus the density-dependent term in the EoM, eq.~\eqref{eq:sym EoM}, have an inverse proportionality to $\Msym^2$, meaning a large change in density has less impact on the field profile. 

The approximation of symmetron screening surfaces as isodensity contours matching $\rhoSSB$ is a natural consequence of the highly sensitive relationship between the screened region and the symmetry-breaking scale. These approximate solutions are shown as grey dashed-dotted lines on the symmetron plots of figure~\ref{fig:3. Partial Screening}. For the most part, this approximation does a good job of calculating the true value of $r_s$, especially in the small $\Msym$, small $\Lc$ regime. Outside of this region of parameter space, the approximation always overestimates the value of $r_s$ when compared to the true value. This overestimation is especially prevalent in the large $\Lc$, large $\Msym$ regime.

\subsection{Amounts of Galactic Mass Screened}
\label{subsec:massscreened}

\begin{figure}[tbp]
\centering 
\includegraphics[width=\textwidth, trim=0 0 0 0, clip]{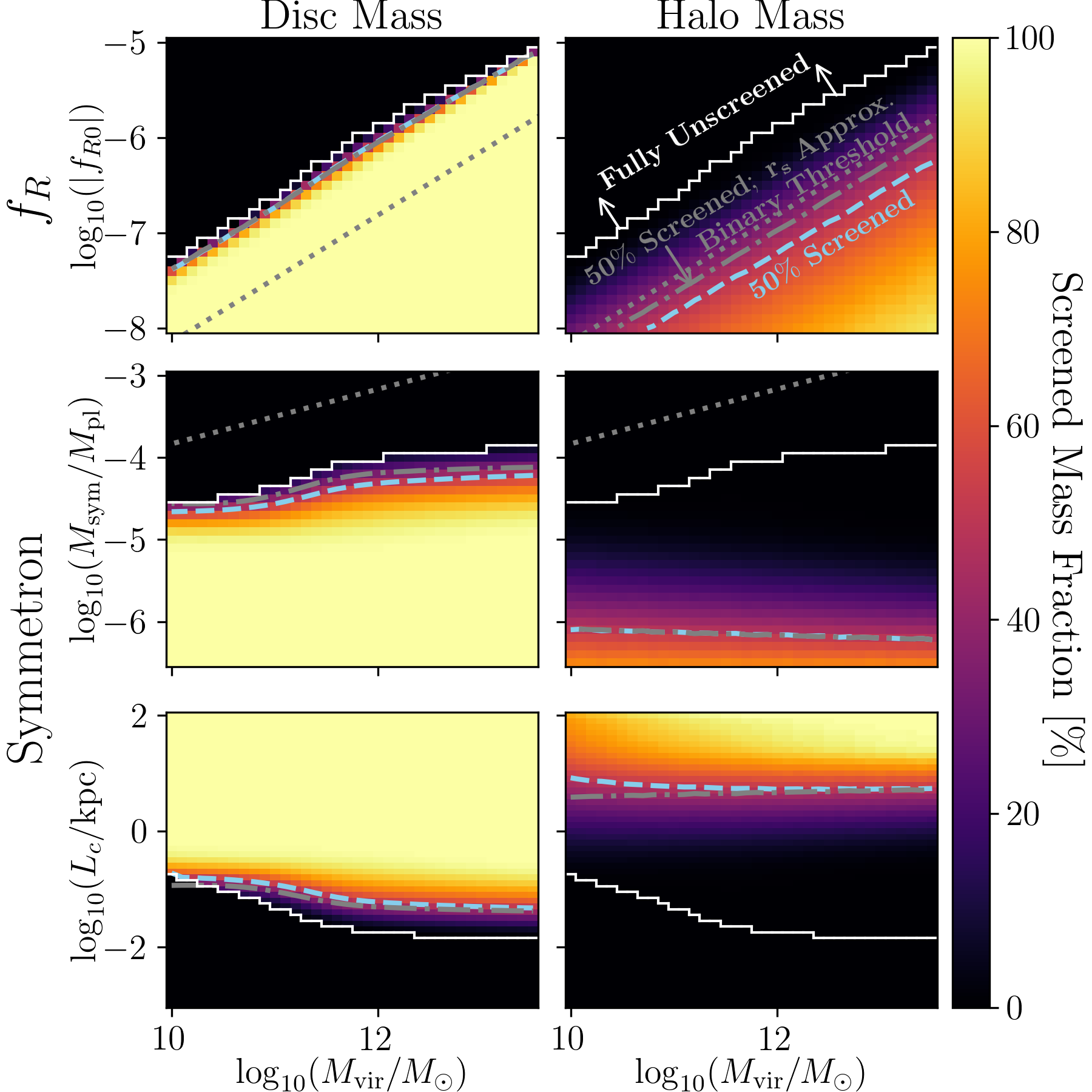}
\hfill
\caption{\label{fig:4. Mass Screened} Fraction of disc mass (\textbf{left}) and dark matter (\textbf{right}) screened, for a range of $\fofR$ (\textbf{top}) and symmetron parameters, with fixed $\Lc=10^{-1}\,\mathrm{kpc}$ (\textbf{middle}) and with fixed $\Msym=10^{-4.5}\,\Mpl$ (\textbf{bottom}). Sky blue dashed contours show where 50\% of mass is screened. Grey dashed-dotted contours indicate where 50\% of mass is screened when $r_s$ is approximated: by numerically integrating eq.~\eqref{eq:chi defn} for the $\fofR$ critical potential, or by finding isodensity contours equal to $\rhoSSB$ for the symmetron model. Grey dotted lines correspond to the binary screening condition; solutions below this line would be considered `fully screened'. Solid white lines separate the fully unscreened and partially screened solutions. }
\end{figure}

One of the main results of our analysis is contained within figure~\ref{fig:4. Mass Screened}, where we show the percentage of mass contained within the screening surface over a range of galactic and theory parameters. The left column represents the disc mass and the right column is the halo mass. In the symmetron plots, we vary one parameter with the other parameter fixed. This figure allows us to study how our system transitions from screened to unscreened across the parameter space, which we can compare with the transition from screened to unscreened given by the binary screening condition, eq.~\eqref{eq:fR crit pot} in the $\fofR$ case and eq.~\eqref{eq:sym crit pot} in the symmetron case. As a proxy for the `true' answer, we include a dashed contour line to show the parameter boundary where 50\% of the mass is screened in each case\footnote{There is ultimately a degree of arbitrariness in our choice of 50\% as a proxy. It intuitively feels the fairest way to divide the spectrum of solutions in two, but one could instead argue for a lower (higher) threshold such that the partially screened solutions are preferentially grouped with the fully screened (unscreened) solutions.}. This line can then be directly compared with the binary screening condition boundary, which we represent with a dotted line in the $\fofR$ and symmetron $\Msym$ plots respectively (not included in the $\Lc$ plot since the symmetron critical potential is a function of $\Msym$ only).

All the sub-figures contained in figure~\ref{fig:4. Mass Screened} exhibit some similar behaviours. The galactic disc component transitions sharply from being fully screened to fully unscreened, with the two regimes separated in the parameter space of galaxy and model parameters by only a narrow band in which the disc is partially screened. In contrast, the fraction of screened dark matter varies gradually, due to the significant dynamical range of the halo, which spans several orders of magnitude in density. Another notable feature is the discrepancy between the regions considered fully screened and unscreened by the binary condition and the region where half of the mass is screened. This is explored further throughout this section. In all cases, the $r_s$ approximations (dashed-dotted lines) are closer to the true 50\% screened solution than the binary screening condition.

Focusing first on the $\fofR$ case, we see that the $\fofR$ binary screening condition, eq.~\eqref{eq:fR crit pot}, does not correctly characterise the screening of the galactic disc. Specifically, the region of parameter space considered `fully unscreened' by the binary condition (above the dotted line in the top left panel) includes roughly an order of magnitude in which the galactic disc is actually entirely screened. This ultimately means that searching for signatures of screening theories in `fully unscreened' galaxies may lead to false negative results since the visible stellar and gas components are fully screened. For example, when looking at stellar-gas offsets, ref.~\cite{Desmond+2018Offsets, Desmond+2019}, in this region of parameter space the disc is fully screened and there is therefore no displacement between the stellar and gas components. However, using eq.~\eqref{eq:fR crit pot} as a binary screening threshold would lead one to assume that the disc is unscreened and thus incorrectly predict the presence of an offset. In practice, this could entail relaxing \hl{the current tightest constraints. }

By contrast with the disc, for the halo the binary screening condition kicks in at a point where the halo is only partially screened, so there will be haloes (below the dotted line in the top right panel) which will be classed as screened but actually carry a substantial unscreened component in their outskirts, as much as 60-70\% of the total halo mass. Similarly, there will be haloes (above the dotted line) which will be classed as unscreened but actually have screened central regions, comprising as much as 30-40\% of the total halo mass. This is the fundamental flaw in the binary screening condition: it assumes a step change in a quantity which actually changes gradually over the course of several orders of magnitude in parameter space.
In summary, the $\fofR$ binary screening condition underestimates unscreened mass when it classifies an object as screened, and it underestimates screened mass when it classifies an object as unscreened.

The various panels in figure~\ref{fig:4. Mass Screened} also include dashed-dotted lines to indicate when 50\% of the mass is screened when $r_s$ is derived by approximate means. In the $\fofR$ case (top row), this is when $r_s$ is derived by numerically integrating eq.~\eqref{eq:chi defn} for a spherically averaged version of our galaxy density profile. This approximation solution yields results closer to the exact solution than the binary screening condition. For the galactic disc (top left panel), this approximation contour is almost identical to the equivalent numerical solution contour, and the difference in the proportion of mass screened between the two calculations is no more than 5\% across the range of parameters we consider. However, as we saw in figure~\ref{fig:3. Partial Screening}, the true and approximate solutions for $r_s$ can have vastly different shapes and the spherical nature of the approximate screening surface results in an underestimation of $r_s$ along the galactic disc and an overestimation along the perpendicular axis. 

For the dark matter component (top right panel), the approximate value of $r_s$, and therefore the percentage of mass screened, is very close to the actual value for low values of $r_s$. 
However, as $r_s$ becomes larger, the approximate solution begins to overestimate $r_s$ compared to the actual solution, as previously seen in figure~\ref{fig:3. Partial Screening}.
\hl{This overestimation occurs in the limit of large $\Mvir$ and small $\fRz$, and emerges because, in this regime, the Compton wavelength becomes smaller than the extent of the galaxy. Given that the approximation assumes a negligible scalar mass and thus an infinite Compton wavelength, it considers mass beyond a Compton wavelength (which does not significantly impact the field). The biggest discrepancy between the approximate and true screening surfaces, located in the lower right of the $\fofR$ parameter space of figure~\ref{fig:4. Mass Screened}, results in the estimated proportion of screened dark matter being as much as 20\% larger than the actual solution.}
In other words, if the 50\% threshold were increased to, e.g., 80\%, then the two lines would be further separated. Ultimately this will result in the predicted fifth forces sourced by the galaxy being larger than in the true case, but not to the same extent as the binary screening condition would predict.

For the symmetron model (middle and bottom rows), the binary screening condition is much less accurate. As discussed in section~\ref{ssec:symmetron}, the condition depends only on $\Msym$, but even in the middle row, with varying $\Msym$ and fixed $\Lc$, this estimate of screening disagrees with our solution by several orders of magnitude. Looking at the row with varying $\Lc$ and fixed $\Msym$, it is clear that the location of the screening surface is a strong function of $\Lc$. Therefore, this simplified version of the symmetron binary screening condition should be avoided.

Instead, a more useful approximation for screening is obtained by assuming the screening surface is coincident with the surface where the total density equals $\rhoSSB$. The parameter space that results in 50\% of each mass component being enclosed by this approximate screening surface is shown in the symmetron panels of figure~\ref{fig:4. Mass Screened} as a dashed-dotted contour. As discussed in section~\ref{ssec: partial screening}, this approximation will never underestimate the value of $r_s$, and will overestimate $r_s$ for large values of $\Msym$ or $\Lc$. Looking first at the disc mass, in both cases (varying $\Msym$ and varying $\Lc$) the approximation contour follows a similar behaviour to the true contour, but somewhat overestimates the amount of disc mass screened. Notably, the distance in parameter space between the true and approximate 50\% screened contours is reduced in the case where $\Lc$ is smaller (towards the right of the bottom-left panel with varying $\Lc$ and fixed $\Msym$). For the dark matter component, we see that the approximate and true 50\% screening lines are indistinguishable in the middle right panel with varying $\Msym$ and fixed $\Lc$. This is due to the fact that $\Lc$ is fixed at a low value of $10^{-1}\,\mathrm{kpc}$, and so to get a large screening radius that screens 50\% of the halo we need a small value of $\Msym$, thus falling into the regime where this approximation works best (small $\Msym$, small $\Lc$). In the bottom right panel with varying $\Lc$ and fixed $\Msym$, the 50\% line is a good match for larger galaxies, but for smaller galaxies the approximation overestimates the amount of mass screened. As length scales get smaller they become comparable with $\Lc$ and the field can not respond to the density changes quickly enough to screen exactly along the isodensity contour. In summary, the approximation does not account for the full effects of varying $\Msym$ and $\Lc$, so it falters in the regimes in which these effects are strongest.

One noteworthy caveat to this discussion is that we do not consider any environmental effects, e.g., the presence of over-/underdensities within a few Compton wavelengths of our target galaxies. In general, these would serve to vary the location of the screening surface of a given galaxy from that calculated in isolation. \hl{We have done some preliminary tests in this regard, and find the conclusions to this section are robust. We experimented with mimicking overdense environments by embedding our galaxies in large overdense spheres of constant density. As expected, the additional overdensity leads to the galaxies being screened more easily. However, we find that either the true and approximate solutions for the screening surface shift equally or the gulf widens. However, a full reanalysis with proper treatment of environmental effects is required to make a clear, quantitative conclusion.}

\subsection{Milky Way Observables} 
\label{subsec:milkyway}

\begin{figure}[htp]
    \centering
    \includegraphics[trim={0 0 0 0}, clip, width=\textwidth]{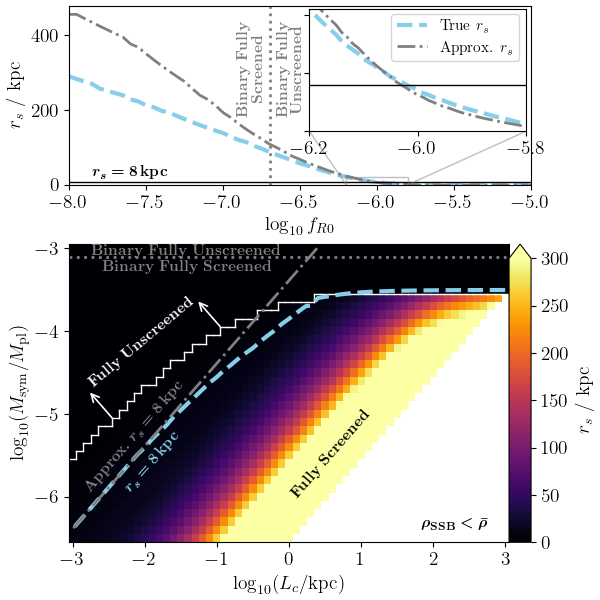}
    \caption{\label{fig:5. MW screening} Positions where non-spherical screening surface intersects the disc plane of a Milky Way-like galaxy over $\fofR$ (\textbf{top}) and symmetron (\textbf{bottom}) parameter space. In the $\fofR$ plot, the dashed line shows the true $r_s$ value, whereas the dashed-dotted line shows the approximate $r_s$ values, derived by numerically integrating eq.~\eqref{eq:chi defn} for the $\fofR$ critical potential. In the symmetron plots, the colourmap shows the true $r_s$ value, the dashed line indicates where $r_s$ is equal to the Milky Way centre-Solar System separation of 8\,kpc, and the dashed-dotted line shows where the approximate $r_s$, derived by finding isodensity contours equal to $\rhoSSB$, is equal to the Milky Way centre-Solar System separation. Dotted lines indicate the binary screening condition. In the symmetron plot, a solid white line separates the fully unscreened and partially screened solutions, and the white region shows where SSB does not occur, i.e. the cosmic background density, $\bar{\rho}$, is greater than $\rhoSSB$.}
\end{figure}

Figure~\ref{fig:5. MW screening} shows the derived screening radius along the plane of the galactic disc, over the entire model parameter spaces, for a Milky Way-like galaxy with $\Mvir = 1.5\times 10^{12}M_\odot$ \cite{McMillan2017}. The approximate distance from the Milky Way centre to the Solar System (8\,kpc) is marked by the dashed lines. The dashed-dotted lines show the screening radius derived via approximate methods, namely, numerically integrating eq.~\eqref{eq:chi defn} given the $\fofR$ critical potential, eq.~\eqref{eq:fR crit pot}, or defining the screening surface as the isodensity contour equal to $\rhoSSB$ in the symmetron case. The dotted lines represent the binary screening boundary, separating the parameters that cause solutions to be considered `fully screened' and `fully unscreened'. Specifically, all $|\fRz|$ values less than $10^{-6.7}$ and all $\Msym$ values less than $10^{-3.1}\,\Mpl$ are considered `fully screened' by the binary screening conditions. \hl{In the $\fofR$ plot, we again see the approximate $r_s$ begins to overestimate the true solutions when the Compton wavelength becomes comparable to the extent of the galaxy, at $\fRz \lesssim 10^{-7}$.}

Stringent constraints based on tests of celestial gravitational dynamics require that the Milky Way screening radius extends to at least 8\,kpc, encompassing the entire Solar System \cite{Adelberger:2009zz}. Figure~\ref{fig:5. MW screening} allows us to determine the parameter space permitted by such constraints, namely $|\fRz| \lesssim 10^{-6.0}$ and the portion of the symmetron parameter space satisfying $\Msym \lesssim 10^{-3.6}\Mpl$ and to the right of the dashed line on the lower panel. There are a few caveats to these constraints. Firstly, they only apply to this particular galaxy model and choice of mass. We do not expect the Milky Way to be a maximally-typical galaxy but for consistency with our previous analysis, we have modelled it as such. In addition, there remains significant empirical uncertainty regarding the total mass of the Milky Way halo \cite{Callingham+2019MWMass}, which would have a non-negligible effect on the bounds we obtain. We have also neglected any environmental effects, such as the influence of the Local Group and the galaxy's wider cosmographic context.

Given the strict constraint on the minimum value of the Milky Way's screening radius, the question arises as to whether there can still be observable effects in the periphery of the galactic disc. For this galaxy, the disc component becomes subdominant along the disc plane after roughly 14\,kpc. Looking at either panel of figure~\ref{fig:5. MW screening}, between the $r_s=8$\,kpc contour and the parameter space where $r_s=14$\,kpc, there is only a small slice of parameter space in which the Solar System is screened but the bulk of the galactic disc is not. This means it would require a great deal of fortune to be in a universe adopting these parameter values and allowing us to explain observable effects in the disc via fifth forces. 

Nevertheless, there remains several orders of magnitude of unconstrained parameter space in which the screening radius lies between the edge of the galactic disc and the virial radius of $\sim$\,300\,kpc. As such, observable consequences of scalar fields with screening mechanisms located in the Milky Way's dark matter halo, are entirely plausible. For example, the kinematics of halo tracers or the detection of equivalence-principle-violating asymmetries in the leading and trailing tails of stellar streams, from tidally disrupted dwarf galaxies in the dark matter halo, as in refs.~\cite{Naik+2020Streams, KesdenKamionkowski2006Streams1, KesdenKamionkowski2006Streams2, Keselman+2009Streams}. 

An additional feature in figure~\ref{fig:5. MW screening} worth mentioning is the turn-off in the top right of the symmetron plot. Away from this region, lines of equal screening radius are diagonals across parameter space. These diagonals correspond with lines of equal $\rhoSSB$ ($\rhoSSB \propto \Msym/\Lc^2$; cf. eq.~\eqref{eq:sym rho_SSB}), and as we saw in figure.~\ref{fig:3. Partial Screening}, screening surfaces closely match the $\rho=\rhoSSB$ surfaces in the low $\Msym$, low $\Lc$ regime. The turn-off is another symptom of the breakdown of this behaviour in the large $\Msym$, large $\Lc$ limit.


\section{Conclusions}
\label{sec:conclusion}

In this work, we have investigated the effects of scalar modifications of gravity on galactic scales, for both chameleon $\fofR$ and symmetron screening theories, with a particular focus on how screening differs between full numerical solutions and common approximations used to derive constraints. To determine the screened region, we solved the quasi-static scalar field EoM numerically, given an input density described by a maximally-typical galaxy consisting of a double-exponential disc profile and a dark matter NFW halo. We developed new and robust screening conditions to compute the position of the screening surface based on the scalar field values, eqs.~\eqref{eq:screening conditions}.

We examined the shape of our calculated screening surfaces. We find that screening surfaces in the halo-dominated region follow the spherical isodensity contours. In the disc-dominated region, $\fofR$ screening surfaces almost follow the isodensity contours, but are overall more spherical than the disc profile. Symmetron screening surfaces closely follow isodensity contours in the disc-dominated region, except in the large $\Msym$, large $\Lc$ limit where the individual microphysics of each parameter becomes important.

The simplest approximation we explore is the binary screening condition, ref.~\cite{Cabre+2012}, where if a galaxy has a virial potential, $\Phi_\mathrm{vir}=G\Mvir/\Rvir$, greater/smaller than the critical potential then it is classified as fully screened/unscreened, ignoring the effects of partial screening. We note that the symmetron binary screening condition, eq.~\eqref{eq:sym crit pot}, is not suitable as a proxy for screening. This is because the approximations required to derive the critical potential, eq.~\eqref{eq:chi defn}, are incompatible with the symmetron theory\hl{, namely the assumptions that the field can be linearised, the scalar mass is negligible and $\beta(\phi)$ is constant.}

We explored the effectiveness of the binary screening condition as a proxy to $\fofR$ screening, for observables in the visible disc and dark matter halo. We found that the binary screening condition drastically underestimates the amount of parameter space in which the galactic disc is fully screened. In this fully screened region of parameter space, using the binary screening condition would erroneously define the galaxy as `fully unscreened', i.e. it would lead one to predict an observable signature that would not occur. It follows that constraints set using the binary screening condition in this contested region of parameter space would be tighter than if one numerically computed the true screening surface. 

We also compare our true solutions against approximate screening surfaces. In the $\fofR$ case, $r_s$ can be approximated by numerically solving the critical potential, eq.~\eqref{eq:chi defn}, for a spherically averaged galactic density. In this case, the resulting approximate screening surfaces are spherical. For the symmetron model, it is possible to approximate $r_s$ as the location where the density is equal to the symmetron breaking scale, i.e. $\rho(r=r_s)=\rhoSSB$. 

For the halo, we found the binary screening condition sits in the region of parameter space where the haloes are partially screened. This means objects classified as ``screened'' actually have 60-70\% unscreened mass, while objects classified as ``unscreened'' actually have 30-40\% screened mass. This is the fundamental flaw of the binary screening condition, the fact that the binary screening condition assumes a discrete step change between fully screened and unscreened, but actually the degree of screening varies gradually over many orders of magnitude in parameter space.

Direct approximation of $r_s$ allows for partial screening solutions and is overall a closer approximation to the true screening solutions than the binary screening condition. In regions where the disc component is not negligible, the spherical approximate screening surface will underestimate $r_s$ along the disc and overestimate $r_s$ perpendicular to the disc. As a result of this, we might expect to see an observable feature at a location along the plane of the disc, but in fact, the location is within the true screening surface and therefore there is no observable feature. In addition, this approximation overestimates $r_s$ for small values of $|\fRz|$, \hl{corresponding to smaller Compton wavelengths. This is a result of the approximation assuming a negligible scalar mass, and thus being influenced by additional mass that the field does not respond as strongly to. The effect is amplified in the large $\Mvir$, small $\fRz$ limit, where the extent of the galaxy is largest and the Compton wavelength is smallest.}

We have shown that defining the symmetron $r_s$ as the radius at which the total density equals $\rhoSSB$ is a useful approximation for a large portion of the parameter space. However, in the large $\Msym$, large $\Lc$ limit, varying the two parameters separately has consequences for the scalar field profile that are not captured purely through a dependence on $\rhoSSB$ and the approximation breaks down. In general, the approximate screening surface will never underestimate $r_s$ and the amount of overestimation depends on how large $\Msym$ and $\Lc$ are. In smaller galaxies, the approximation may break down if the galactic scale length becomes comparable to $\Lc$. 

Finally, we explored the possibility of finding observables in a Milky Way-like galaxy. Assuming that the Solar System is screened by the Milky Way, we found that (for our isolated test Milky-Way galaxy) $|\fRz|$ must be less than $10^{-6.0}$ and $\Msym$ must be less than $10^{-3.6}\,\Mpl$ with $\Lc$ being to the right of the dashed line in the bottom panel of figure~\ref{fig:5. MW screening}. We saw that there is only a small slice of model parameter space in which there could be observable effects in the galactic disc. This suggests we would have to be rather fortunate to exist in a universe where we could detect the modifications to GR considered in this work within our own galactic disc --- however, searches for observables in the halo remain promising, thanks in part to the large dynamical range of the halo density.

As cosmological surveys continue to improve the precision of their measurements it is important that theoretical predictions continue to improve in accuracy alongside them. It is in this spirit that we have presented, in this paper, new definitions of screening for both the $\fofR$ chameleon model and the symmetron model of screening. We have shown that it is important to differentiate between situations where the galactic disc is screened, but much of the dark matter halo is unscreened, and situations where the majority of the total mass of the galaxy is screened. We have shown that some approximations currently used in the literature could lead to order-of-magnitude errors in the constraints placed on model parameters from galactic observations. We intend to present a reanalysis of current constraints in the near future.

\appendix

\section{\texorpdfstring{$\chi$}{chi} Parameter Derivation}
\label{sec:app chi parameter derivation}

This appendix provides a complete derivation of the critical potential, eq.~\eqref{eq:chi defn}, introduced in section~\ref{ssec:astro screening approximations}. This prevalent approximation to screening comes from solving the general EoM eq.~\eqref{eq:general EoM} for a static spherically-symmetric source, $\delta\rho(r)$, of radius $R$, in a constant density background, $\bar{\rho}$. This method can be applied to scenarios such as a star embedded in the galactic medium or a dark matter halo embedded in the cosmological background.

Far from the object, the field settles into the minimum of the effective potential, i.e. $\phi \rightarrow \bar{\phi} \equiv \phi_\mathrm{min}(\bar{\rho})$. Near the object, if the field is able to reach the minimum of the potential inside the object then $\nabla^2\phi=0$, meaning $\phi=\phi_\mathrm{min}(\rho)$, and the fifth force is screened due to the vanishing field gradients. Otherwise, the EoM needs to be solved, which can be done if we assume the object is static and spherically symmetric, and therefore $\Box \rightarrow \nabla^2(r)$. If the field remains close to the background value, $\bar{\phi}$, we linearise as $\phi = \bar{\phi} + \delta\phi$, \hl{and assume constant $\beta(\phi)$ throughout,} resulting in the EoM
\begin{equation} \label{eq:general exterior EoM}
    \nabla^2\phi = m_0^2\delta\phi + \frac{8\pi G}{c^2}\beta(\bar{\phi})\delta\rho(r) \,,
\end{equation}
where $m_0^2 = V''(\bar{\phi})$. If we assume that $\phi$ is sufficient light in the background environment, such that \hl{$m_0 R \ll 1$}, we find that we can consistently neglect the first term on the RHS of our linearised EoM. In summary, well inside the object, the field is at the minimum of the effective potential, then, near $r_s$, it begins to roll towards the background value.

Now we can solve the EoM outside the screened region by integrating eq.~\eqref{eq:general exterior EoM} (with the assumptions laid out in the previous paragraph) to get
\begin{equation}
    \frac{d\phi}{dr} = \frac{2G\beta(\bar{\phi})}{c^2} \frac{\mathcal{M}(r) - \mathcal{M}(r_s)}{r^2} \,,
\end{equation}
where $\mathcal{M}(r)=\int^r_0 4\pi r'^2\delta\rho(r')dr'$. Then we determine the screening radius by integrating again, this time from $r_s$, where $\phi = \phi_s \approx \phi_\mathrm{min}(\rho)$, to infinity, where the field assumes the background value $\phi=\bar{\phi}$,
\begin{equation}
    \bar{\phi} - \phi_s = \frac{2G\beta(\bar{\phi})}{c^2} \left( \frac{\mathcal{M}(r_s)}{r_s} 
    + \int^\infty_{r_s}\frac{\mathcal{M}(r)}{r^2} dr \right)\,.
\end{equation}
Finally, integrating by parts and using our expression for $\mathcal{M}(r)$, we find the expression
\begin{equation} \label{eq:app chi defn}
    \chi \equiv \frac{\bar{\phi}}{2\beta(\bar{\phi})} = \frac{4\pi G}{c^2} \int^\infty_{r_s} r\delta\rho(r)dr \,,
\end{equation}
where the $\phi_s$ term has been dropped since it is assumed the screening mechanism acts to force the field to smaller values inside high-density objects, thus $\phi_s \ll \bar{\phi}$. 

\section{Galaxy Pipeline}
\label{sec:App. galaxy pipeline}

In this appendix we outline our procedure for deriving all five of our galaxy parameters ($\rhonfw,\ \rnfw,\ \Sigmadisc,\ \Rdisc$ and $\zdisc$), introduced in section~\ref{ssec:galaxy model} via eqs.~\eqref{eq:NFW profile} and \eqref{eq:disc profile}, in terms of one input parameter, $\Mvir$. To achieve this we take advantage of several empirical scaling relations. Note that all non-analytic relations come with a related dispersion. As discussed in the main text, section~\ref{ssec:galaxy model}, we ignore the scatter about the scaling relations, and simply adopt median values in each case.

\subsection{NFW Parameters} \label{ssec:NFW parameter pipeline}

On the dark matter side, an NFW profile eq.~\eqref{eq:NFW profile} is defined in terms of one of two sets of input parameters, $\rhonfw$ and $\rnfw$, or $\Mvir$ and $\cvir$. We choose the latter since we already have one of the required input parameters, $\Mvir$, equivalent to the commonly used $M_{200}$. As such, we only need to employ a single empirical relation between the virial mass and concentration parameter. For this, we use equation 8 of ref.~\cite{DuttonMaccio2014},
\begin{equation} \label{eq:virial mass concentration relation}
    \log_{10} \cvir = 0.905 - 0.101\log_{10}\left(\frac{\Mvir}{10^{12}h^{-1}M_\odot}\right) \,,
\end{equation}
where $h$ is the Hubble parameter, taken to be $0.7$. This relation reflects the well-known result from $\Lambda$CDM simulations that more massive dark matter haloes are less centrally concentrated. The intrinsic scatter is given as log-normal with $\sigma_{\log_{10} \cvir}=0.11$, (see end of \textsection 4.1 in ref.~\cite{DuttonMaccio2014}).

At this point, $\rhonfw$ and $\rnfw$ can be obtained from $\Mvir$ and $\cvir$ from simultaneous solution of
\begin{equation}
    \Mvir = \frac{4}{3} \pi \Delta \rho_\mathrm{crit} \Rvir^3 \,,
\end{equation}
\begin{equation}
    \Rvir = \cvir\rnfw \,,
\end{equation}
\begin{equation} \label{eq:virial mass}
    \Mvir  = 4\pi \rhonfw \rnfw^3 \left[\ln(1+\cvir) - \frac{\cvir}{1+\cvir}\right] \,,
\end{equation}
which derive respectively from the definition of the virial mass, the definition of the concentration, and the relation for the enclosed mass at a given radius under the NFW profile.

\subsection{Galactic Disc Parameters} \label{ssec:galactic disc parameter pipeline}

To derive the three parameters ($\Sigmadisc$, $\Rdisc$ and $\zdisc$) needed to construct the galactic disc profile, eq.~\eqref{eq:disc profile}, we must pass from the virial mass to the stellar mass (contributing to the bulk of the galactic disc). To go from halo to disc mass, we employ the empirical stellar-halo mass relation (SHMR), namely eq.~2 in ref.~\cite{Moster2013}
\begin{equation}
    \Mdisc = 2N\Mhalo\left[\left(\frac{\Mhalo}{M_1}\right)^{-\beta} +\left(\frac{\Mhalo}{M_1}\right)^\gamma\right]^{-1} \,,
\end{equation}
where the four fitting parameters and related errors are in table 3 of ref.~\cite{Moster2013}. 

Now that we have the disc mass, we pass via the half-light radius to the disc scale length, $\Rdisc$. To achieve this, we use the observational relation defined by eq.~18 in ref.~\cite{Shen2003}. Specifically,
\begin{gather}
    \frac{\overline{R}_{hl}}{\mathrm{kpc}} = \gamma \left(\frac{\Mdisc}{M_\odot}\right)^\alpha \left(1+\frac{\Mdisc}{M_0}\right)^{\beta-\alpha} \,,
\end{gather}
where $\overline{R}_{hl}$ denotes the median half-light radius and fitting parameters ($\alpha$, $\beta$, $\gamma$, $M_0$, $\sigma_1$ and $\sigma_2$) take the values from the bottom row of table~1 in ref.~\cite{Shen2003}. Then, by integrating the density, eq.~\eqref{eq:disc profile}, to calculate the mass, and assuming a constant mass-to-light ratio so that we can utilise $\Mdisc(R=\overline{R}_{hl}) = \frac{1}{2}\Mdisc(R\rightarrow\infty)$, we arrive at the relation 
\begin{equation}
    \Rdisc = -\overline{R}_{hl}\left(1+W_{-1}\left(-\frac{1}{2e}\right)\right) \approx 0.595824 \overline{R}_{hl} \,,
\end{equation}
where $W_{-1}(x)$ is the $n=-1$ solution to the Lambert $W_n$ function.

The disc mass can be determined by integrating the density profile, eq.~\eqref{eq:disc profile}, 
\begin{equation}
    \Mdisc(R\rightarrow\infty) = 2\pi\Sigmadisc \Rdisc^2 \,,
\end{equation}
at which point it is easily reversed and combined with our value of $\Rdisc$ to calculate $\Sigmadisc$.

Finally, we employ one last observational relation to determine the disc scale height, $\zdisc$, in terms of the disc scale radius $\Rdisc$. For this we utilise eq.~1 of ref.~\cite{Bershady2010},
\begin{equation}
    \log_{10}\left(\frac{\Rdisc}{\zdisc}\right) = 0.367\log_{10}\left(\frac{\Rdisc}{\mathrm{kpc}}\right) + 0.708 \pm 0.095 \,.
\end{equation}

With this, we have defined all five density and length scales ($\rhonfw,\ \rnfw,\ \Sigmadisc,\ \Rdisc$ and $\zdisc$) in terms of a single input, $\Mvir$.

\acknowledgments
C.B. is supported by a Research Leadership Award from the Leverhulme Trust and by the STFC under grant ST/T000732/1. B.M. is supported by an STFC studentship. A.P.N. is supported by an Early Career Fellowship from the Leverhulme Trust. For the purpose of open access, the authors have applied a Creative Commons Attribution (CC BY) licence to any Author Accepted Manuscript version arising.

\section*{Data Access Statement}

The source code and plotting scripts are publicly available at
\href{https://github.com/Bradley-March/GalacticScreeningConditions/}{github.com/Bradley-March} \href{https://github.com/Bradley-March/GalacticScreeningConditions}{GalacticScreeningConditions}. Scalar field solutions
can be made available on reasonable request to the authors.


\bibliographystyle{JHEP}
\bibliography{library}

\providecommand{\href}[2]{#2}\begingroup\raggedright\begin{thebibliography}{10}

\bibitem{Adelberger2003}
E.G.~{Adelberger}, B.R.~{Heckel} and A.E.~{Nelson}, \emph{{Tests of the
  Gravitational Inverse-Square Law}},
  \href{https://doi.org/10.1146/annurev.nucl.53.041002.110503}{\emph{Annual
  Review of Nuclear and Particle Science} {\bfseries 53} (2003) 77}
  [\href{https://arxiv.org/abs/hep-ph/0307284}{{\ttfamily hep-ph/0307284}}].

\bibitem{Joyce:2014kja}
A.~Joyce, B.~Jain, J.~Khoury and M.~Trodden, \emph{{Beyond the Cosmological
  Standard Model}},
  \href{https://doi.org/10.1016/j.physrep.2014.12.002}{\emph{Phys. Rept.}
  {\bfseries 568} (2015) 1} [\href{https://arxiv.org/abs/1407.0059}{{\ttfamily
  1407.0059}}].

\bibitem{Koyama:2015vza}
K.~Koyama, \emph{{Cosmological Tests of Modified Gravity}},
  \href{https://doi.org/10.1088/0034-4885/79/4/046902}{\emph{Rept. Prog. Phys.}
  {\bfseries 79} (2016) 046902}
  [\href{https://arxiv.org/abs/1504.04623}{{\ttfamily 1504.04623}}].

\bibitem{Ishak:2018his}
M.~Ishak, \emph{{Testing General Relativity in Cosmology}},
  \href{https://doi.org/10.1007/s41114-018-0017-4}{\emph{Living Rev. Rel.}
  {\bfseries 22} (2019) 1} [\href{https://arxiv.org/abs/1806.10122}{{\ttfamily
  1806.10122}}].

\bibitem{BurrageSakstein2018}
C.~{Burrage} and J.~{Sakstein}, \emph{{Tests of chameleon gravity}},
  \href{https://doi.org/10.1007/s41114-018-0011-x}{\emph{Living Reviews in
  Relativity} {\bfseries 21} (2018) 1}
  [\href{https://arxiv.org/abs/1709.09071}{{\ttfamily 1709.09071}}].

\bibitem{Brax:2021wcv}
P.~Brax, S.~Casas, H.~Desmond and B.~Elder, \emph{{Testing Screened Modified
  Gravity}}, \href{https://doi.org/10.3390/universe8010011}{\emph{Universe}
  {\bfseries 8} (2021) 11} [\href{https://arxiv.org/abs/2201.10817}{{\ttfamily
  2201.10817}}].

\bibitem{CANTATA:2021ktz}
E.N.~{Saridakis}, R.~{Lazkoz}, V.~{Salzano}, P.~{Vargas Moniz},
  S.~{Capozziello}, J.~{Beltr{\'a}n Jim{\'e}nez} et~al., \emph{{Modified
  Gravity and Cosmology: An Update by the CANTATA Network}} (May, 2021),
  \href{https://doi.org/10.48550/arXiv.2105.12582}{10.48550/arXiv.2105.12582},
  [\href{https://arxiv.org/abs/2105.12582}{{\ttfamily 2105.12582}}].

\bibitem{Vardanyan:2023jkm}
V.~Vardanyan and D.J.~Bartlett, \emph{{Modeling and Testing Screening
  Mechanisms in the Laboratory and in Space}},
  \href{https://doi.org/10.3390/universe9070340}{\emph{Universe} {\bfseries 9}
  (2023) 340} [\href{https://arxiv.org/abs/2305.18899}{{\ttfamily
  2305.18899}}].

\bibitem{Khoury:2003aq}
J.~Khoury and A.~Weltman, \emph{{Chameleon fields: Awaiting surprises for tests
  of gravity in space}},
  \href{https://doi.org/10.1103/PhysRevLett.93.171104}{\emph{Phys. Rev. Lett.}
  {\bfseries 93} (2004) 171104}
  [\href{https://arxiv.org/abs/astro-ph/0309300}{{\ttfamily
  astro-ph/0309300}}].

\bibitem{KhouryWeltman2004}
J.~{Khoury} and A.~{Weltman}, \emph{{Chameleon cosmology}},
  \href{https://doi.org/10.1103/PhysRevD.69.044026}{\emph{\prd} {\bfseries 69}
  (2004) 044026} [\href{https://arxiv.org/abs/astro-ph/0309411}{{\ttfamily
  astro-ph/0309411}}].

\bibitem{HinterbichlerKhoury2010}
K.~{Hinterbichler} and J.~{Khoury}, \emph{{Screening Long-Range Forces through
  Local Symmetry Restoration}},
  \href{https://doi.org/10.1103/PhysRevLett.104.231301}{\emph{\prl} {\bfseries
  104} (2010) 231301} [\href{https://arxiv.org/abs/1001.4525}{{\ttfamily
  1001.4525}}].

\bibitem{Hinterbichler:2011ca}
K.~Hinterbichler, J.~Khoury, A.~Levy and A.~Matas, \emph{{Symmetron
  Cosmology}}, \href{https://doi.org/10.1103/PhysRevD.84.103521}{\emph{Phys.
  Rev. D} {\bfseries 84} (2011) 103521}
  [\href{https://arxiv.org/abs/1107.2112}{{\ttfamily 1107.2112}}].

\bibitem{Mota:2006fz}
D.F.~Mota and D.J.~Shaw, \emph{{Evading Equivalence Principle Violations,
  Cosmological and other Experimental Constraints in Scalar Field Theories with
  a Strong Coupling to Matter}},
  \href{https://doi.org/10.1103/PhysRevD.75.063501}{\emph{Phys. Rev. D}
  {\bfseries 75} (2007) 063501}
  [\href{https://arxiv.org/abs/hep-ph/0608078}{{\ttfamily hep-ph/0608078}}].

\bibitem{Dehnen:1992rr}
H.~Dehnen, H.~Frommert and F.~Ghaboussi, \emph{{Higgs field and a new scalar -
  tensor theory of gravity}},
  \href{https://doi.org/10.1007/BF00674344}{\emph{Int. J. Theor. Phys.}
  {\bfseries 31} (1992) 109}.

\bibitem{Gessner:1992flm}
E.~Gessner, \emph{{A new scalar tensor theory for gravity and the flat rotation
  curves of spiral galaxies}},
  \href{https://doi.org/10.1007/BF00645239}{\emph{Astrophys. Space Sci.}
  {\bfseries 196} (1992) 29}.

\bibitem{Damour:1994zq}
T.~Damour and A.M.~Polyakov, \emph{{The String dilaton and a least coupling
  principle}}, \href{https://doi.org/10.1016/0550-3213(94)90143-0}{\emph{Nucl.
  Phys. B} {\bfseries 423} (1994) 532}
  [\href{https://arxiv.org/abs/hep-th/9401069}{{\ttfamily hep-th/9401069}}].

\bibitem{Pietroni:2005pv}
M.~Pietroni, \emph{{Dark energy condensation}},
  \href{https://doi.org/10.1103/PhysRevD.72.043535}{\emph{Phys. Rev. D}
  {\bfseries 72} (2005) 043535}
  [\href{https://arxiv.org/abs/astro-ph/0505615}{{\ttfamily
  astro-ph/0505615}}].

\bibitem{Olive:2007aj}
K.A.~Olive and M.~Pospelov, \emph{{Environmental dependence of masses and
  coupling constants}},
  \href{https://doi.org/10.1103/PhysRevD.77.043524}{\emph{Phys. Rev. D}
  {\bfseries 77} (2008) 043524}
  [\href{https://arxiv.org/abs/0709.3825}{{\ttfamily 0709.3825}}].

\bibitem{DGP2000}
G.~{Dvali}, G.~{Gabadadze} and M.~{Porrati}, \emph{{4D gravity on a brane in 5D
  Minkowski space}},
  \href{https://doi.org/10.1016/S0370-2693(00)00669-9}{\emph{Physics Letters B}
  {\bfseries 485} (2000) 208}
  [\href{https://arxiv.org/abs/hep-th/0005016}{{\ttfamily hep-th/0005016}}].

\bibitem{Vainshtein:1972sx}
A.I.~Vainshtein, \emph{{To the problem of nonvanishing gravitation mass}},
  \href{https://doi.org/10.1016/0370-2693(72)90147-5}{\emph{Phys. Lett. B}
  {\bfseries 39} (1972) 393}.

\bibitem{Hui+2009}
L.~{Hui}, A.~{Nicolis} and C.W.~{Stubbs}, \emph{{Equivalence principle
  implications of modified gravity models}},
  \href{https://doi.org/10.1103/PhysRevD.80.104002}{\emph{\prd} {\bfseries 80}
  (2009) 104002} [\href{https://arxiv.org/abs/0905.2966}{{\ttfamily
  0905.2966}}].

\bibitem{JainVanderPlas2011}
B.~{Jain} and J.~{VanderPlas}, \emph{{Tests of modified gravity with dwarf
  galaxies}}, \href{https://doi.org/10.1088/1475-7516/2011/10/032}{\emph{\jcap}
  {\bfseries 2011} (2011) 032}
  [\href{https://arxiv.org/abs/1106.0065}{{\ttfamily 1106.0065}}].

\bibitem{Bartlett+2021}
D.J.~{Bartlett}, H.~{Desmond} and P.G.~{Ferreira}, \emph{{Calibrating galaxy
  formation effects in galactic tests of fundamental physics}},
  \href{https://doi.org/10.1103/PhysRevD.103.123502}{\emph{\prd} {\bfseries
  103} (2021) 123502} [\href{https://arxiv.org/abs/2103.10356}{{\ttfamily
  2103.10356}}].

\bibitem{Desmond+2018Offsets}
H.~{Desmond}, P.G.~{Ferreira}, G.~{Lavaux} and J.~{Jasche}, \emph{{Fifth force
  constraints from the separation of galaxy mass components}},
  \href{https://doi.org/10.1103/PhysRevD.98.064015}{\emph{\prd} {\bfseries 98}
  (2018) 064015} [\href{https://arxiv.org/abs/1807.01482}{{\ttfamily
  1807.01482}}].

\bibitem{Desmond+2018Warps}
H.~{Desmond}, P.G.~{Ferreira}, G.~{Lavaux} and J.~{Jasche}, \emph{{Fifth force
  constraints from galaxy warps}},
  \href{https://doi.org/10.1103/PhysRevD.98.083010}{\emph{\prd} {\bfseries 98}
  (2018) 083010} [\href{https://arxiv.org/abs/1807.11742}{{\ttfamily
  1807.11742}}].

\bibitem{Desmond+2019}
H.~{Desmond}, P.G.~{Ferreira}, G.~{Lavaux} and J.~{Jasche}, \emph{{The fifth
  force in the local cosmic web}},
  \href{https://doi.org/10.1093/mnrasl/sly221}{\emph{\mnras} {\bfseries 483}
  (2019) L64} [\href{https://arxiv.org/abs/1802.07206}{{\ttfamily
  1802.07206}}].

\bibitem{DesmondFerreira2020}
H.~{Desmond} and P.G.~{Ferreira}, \emph{{Galaxy morphology rules out
  astrophysically relevant Hu-Sawicki f (R ) gravity}},
  \href{https://doi.org/10.1103/PhysRevD.102.104060}{\emph{\prd} {\bfseries
  102} (2020) 104060} [\href{https://arxiv.org/abs/2009.08743}{{\ttfamily
  2009.08743}}].

\bibitem{Naik+2020Streams}
A.P.~{Naik}, N.W.~{Evans}, E.~{Puchwein}, H.~{Zhao} and A.C.~{Davis},
  \emph{{Stellar streams in chameleon gravity}},
  \href{https://doi.org/10.1103/PhysRevD.102.084066}{\emph{\prd} {\bfseries
  102} (2020) 084066} [\href{https://arxiv.org/abs/2002.05738}{{\ttfamily
  2002.05738}}].

\bibitem{Vikram+2013}
V.~{Vikram}, A.~{Cabr{\'e}}, B.~{Jain} and J.T.~{VanderPlas},
  \emph{{Astrophysical tests of modified gravity: the morphology and kinematics
  of dwarf galaxies}},
  \href{https://doi.org/10.1088/1475-7516/2013/08/020}{\emph{\jcap} {\bfseries
  2013} (2013) 020} [\href{https://arxiv.org/abs/1303.0295}{{\ttfamily
  1303.0295}}].

\bibitem{Pedersen:2023ina}
E.M.~Pedersen and C.W.~Stubbs, \emph{{Using Elliptical Galaxy Kinematics to
  Compare of the Strength of Gravity in Cosmological Regions of Differing
  Gravitational Potential -- A First Look}},
  \href{https://arxiv.org/abs/2304.02123}{{\ttfamily 2304.02123}}.

\bibitem{Naik+2018}
A.P.~{Naik}, E.~{Puchwein}, A.-C.~{Davis} and C.~{Arnold}, \emph{{Imprints of
  Chameleon f(R) gravity on Galaxy rotation curves}},
  \href{https://doi.org/10.1093/mnras/sty2199}{\emph{\mnras} {\bfseries 480}
  (2018) 5211} [\href{https://arxiv.org/abs/1805.12221}{{\ttfamily
  1805.12221}}].

\bibitem{Naik+2019}
A.P.~{Naik}, E.~{Puchwein}, A.-C.~{Davis}, D.~{Sijacki} and H.~{Desmond},
  \emph{{Constraints on chameleon f(R)-gravity from galaxy rotation curves of
  the SPARC sample}},
  \href{https://doi.org/10.1093/mnras/stz2131}{\emph{\mnras} {\bfseries 489}
  (2019) 771} [\href{https://arxiv.org/abs/1905.13330}{{\ttfamily
  1905.13330}}].

\bibitem{Vikram+2014}
V.~{Vikram}, J.~{Sakstein}, C.~{Davis} and A.~{Neil}, \emph{{Astrophysical
  Tests of Modified Gravity: Stellar and Gaseous Rotation Curves in Dwarf
  Galaxies}}, \href{https://doi.org/10.48550/arXiv.1407.6044}{\emph{arXiv
  e-prints} (2014) arXiv:1407.6044}
  [\href{https://arxiv.org/abs/1407.6044}{{\ttfamily 1407.6044}}].

\bibitem{LombriserPenarrubia2015}
L.~{Lombriser} and J.~{Pe{\~n}arrubia}, \emph{{How chameleons core dwarfs with
  cusps}}, \href{https://doi.org/10.1103/PhysRevD.91.084022}{\emph{\prd}
  {\bfseries 91} (2015) 084022}
  [\href{https://arxiv.org/abs/1407.7862}{{\ttfamily 1407.7862}}].

\bibitem{Burrage+2017}
C.~{Burrage}, E.J.~{Copeland} and P.~{Millington}, \emph{{Radial acceleration
  relation from symmetron fifth forces}},
  \href{https://doi.org/10.1103/PhysRevD.95.064050}{\emph{\prd} {\bfseries 95}
  (2017) 064050} [\href{https://arxiv.org/abs/1610.07529}{{\ttfamily
  1610.07529}}].

\bibitem{Cataldi+2022}
P.~{Cataldi}, S.~{Pedrosa}, N.~{Padilla}, S.~{Landau}, C.~{Arnold} and B.~{Li},
  \emph{{Fingerprints of modified gravity on galaxies in voids}},
  \href{https://doi.org/10.1093/mnras/stac2122}{\emph{\mnras} {\bfseries 515}
  (2022) 5358} [\href{https://arxiv.org/abs/2207.12917}{{\ttfamily
  2207.12917}}].

\bibitem{O'HareBurrage2018}
C.A.J.~{O'Hare} and C.~{Burrage}, \emph{{Stellar kinematics from the symmetron
  fifth force in the Milky Way disk}},
  \href{https://doi.org/10.1103/PhysRevD.98.064019}{\emph{\prd} {\bfseries 98}
  (2018) 064019} [\href{https://arxiv.org/abs/1805.05226}{{\ttfamily
  1805.05226}}].

\bibitem{HogasMortsell2023}
M.~{H{\"o}g{\^a}s} and E.~{M{\"o}rtsell}, \emph{{Impact of symmetron screening
  on the Hubble tension: New constraints using cosmic distance ladder data}},
  \href{https://doi.org/10.1103/PhysRevD.108.024007}{\emph{\prd} {\bfseries
  108} (2023) 024007} [\href{https://arxiv.org/abs/2303.12827}{{\ttfamily
  2303.12827}}].

\bibitem{Baker+2019}
T.~{Baker}, A.~{Barreira}, H.~{Desmond}, P.~{Ferreira}, B.~{Jain}, K.~{Koyama}
  et~al., \emph{{The Novel Probes Project -- Tests of Gravity on Astrophysical
  Scales}}, \href{https://doi.org/10.48550/arXiv.1908.03430}{\emph{arXiv
  e-prints} (2019) arXiv:1908.03430}
  [\href{https://arxiv.org/abs/1908.03430}{{\ttfamily 1908.03430}}].

\bibitem{SDSS2023}
A.~{Almeida}, S.F.~{Anderson}, M.~{Argudo-Fern{\'a}ndez}, C.~{Badenes},
  K.~{Barger}, J.K.~{Barrera-Ballesteros} et~al., \emph{{The Eighteenth Data
  Release of the Sloan Digital Sky Surveys: Targeting and First Spectra from
  SDSS-V}}, \href{https://doi.org/10.3847/1538-4365/acda98}{\emph{\apjs}
  {\bfseries 267} (2023) 44}
  [\href{https://arxiv.org/abs/2301.07688}{{\ttfamily 2301.07688}}].

\bibitem{2M++2011}
G.~{Lavaux} and M.J.~{Hudson}, \emph{{The 2M++ galaxy redshift catalogue}},
  \href{https://doi.org/10.1111/j.1365-2966.2011.19233.x}{\emph{\mnras}
  {\bfseries 416} (2011) 2840}
  [\href{https://arxiv.org/abs/1105.6107}{{\ttfamily 1105.6107}}].

\bibitem{Desi2023}
{DESI Collaboration}, A.G.~{Adame}, J.~{Aguilar}, S.~{Ahlen}, S.~{Alam},
  G.~{Aldering} et~al., \emph{{The Early Data Release of the Dark Energy
  Spectroscopic Instrument}},
  \href{https://doi.org/10.48550/arXiv.2306.06308}{\emph{arXiv e-prints} (2023)
  arXiv:2306.06308} [\href{https://arxiv.org/abs/2306.06308}{{\ttfamily
  2306.06308}}].

\bibitem{Puchwein+2013MG-GADGET}
E.~{Puchwein}, M.~{Baldi} and V.~{Springel}, \emph{{Modified-Gravity-GADGET: a
  new code for cosmological hydrodynamical simulations of modified gravity
  models}}, \href{https://doi.org/10.1093/mnras/stt1575}{\emph{\mnras}
  {\bfseries 436} (2013) 348}
  [\href{https://arxiv.org/abs/1305.2418}{{\ttfamily 1305.2418}}].

\bibitem{Bose2017(Solver)}
S.~{Bose}, B.~{Li}, A.~{Barreira}, J.-h.~{He}, W.A.~{Hellwing}, K.~{Koyama}
  et~al., \emph{{Speeding up N-body simulations of modified gravity: chameleon
  screening models}},
  \href{https://doi.org/10.1088/1475-7516/2017/02/050}{\emph{\jcap} {\bfseries
  2017} (2017) 050} [\href{https://arxiv.org/abs/1611.09375}{{\ttfamily
  1611.09375}}].

\bibitem{Hui+2009CritPotCalc}
L.~{Hui}, A.~{Nicolis} and C.W.~{Stubbs}, \emph{Equivalence principle
  implications of modified gravity models},
  \href{https://doi.org/10.1103/PhysRevD.80.104002}{\emph{Phys. Rev. D}
  {\bfseries 80} (2009) 104002}.

\bibitem{Davis+2012CritPotCalc}
A.-C.~{Davis}, E.A.~{Lim}, J.~{Sakstein} and D.J.~{Shaw}, \emph{Modified
  gravity makes galaxies brighter},
  \href{https://doi.org/10.1103/PhysRevD.85.123006}{\emph{Phys. Rev. D}
  {\bfseries 85} (2012) 123006}.

\bibitem{Sakstein2013CritPotCalc}
J.~{Sakstein}, \emph{Stellar oscillations in modified gravity},
  \href{https://doi.org/10.1103/PhysRevD.88.124013}{\emph{Phys. Rev. D}
  {\bfseries 88} (2013) 124013}.

\bibitem{BurrageSakstein2016CritPotCalc}
C.~{Burrage} and J.~{Sakstein}, \emph{{A compendium of chameleon constraints}},
  \href{https://doi.org/10.1088/1475-7516/2016/11/045}{\emph{\jcap} {\bfseries
  2016} (2016) 045} [\href{https://arxiv.org/abs/1609.01192}{{\ttfamily
  1609.01192}}].

\bibitem{Arnold+2016}
C.~{Arnold}, V.~{Springel} and E.~{Puchwein}, \emph{{Zoomed cosmological
  simulations of Milky Way-sized haloes in f(R) gravity}},
  \href{https://doi.org/10.1093/mnras/stw1708}{\emph{\mnras} {\bfseries 462}
  (2016) 1530} [\href{https://arxiv.org/abs/1604.06095}{{\ttfamily
  1604.06095}}].

\bibitem{Cabre+2012}
A.~{Cabr{\'e}}, V.~{Vikram}, G.-B.~{Zhao}, B.~{Jain} and K.~{Koyama},
  \emph{{Astrophysical tests of gravity: a screening map of the nearby
  universe}}, \href{https://doi.org/10.1088/1475-7516/2012/07/034}{\emph{\jcap}
  {\bfseries 2012} (2012) 034}
  [\href{https://arxiv.org/abs/1204.6046}{{\ttfamily 1204.6046}}].

\bibitem{Desmond+2018Maps}
H.~{Desmond}, P.G.~{Ferreira}, G.~{Lavaux} and J.~{Jasche},
  \emph{{Reconstructing the gravitational field of the local Universe}},
  \href{https://doi.org/10.1093/mnras/stx3062}{\emph{\mnras} {\bfseries 474}
  (2018) 3152} [\href{https://arxiv.org/abs/1705.02420}{{\ttfamily
  1705.02420}}].

\bibitem{SotiriouFaraoni2010}
T.P.~{Sotiriou} and V.~{Faraoni}, \emph{{f(R) theories of gravity}},
  \href{https://doi.org/10.1103/RevModPhys.82.451}{\emph{Reviews of Modern
  Physics} {\bfseries 82} (2010) 451}
  [\href{https://arxiv.org/abs/0805.1726}{{\ttfamily 0805.1726}}].

\bibitem{Brax+2008}
P.~{Brax}, C.~{van de Bruck}, A.-C.~{Davis} and D.J.~{Shaw}, \emph{{f(R)
  gravity and chameleon theories}},
  \href{https://doi.org/10.1103/PhysRevD.78.104021}{\emph{\prd} {\bfseries 78}
  (2008) 104021} [\href{https://arxiv.org/abs/0806.3415}{{\ttfamily
  0806.3415}}].

\bibitem{Faulkner+2007f(R)ChameleonModels}
T.~{Faulkner}, M.~{Tegmark}, E.F.~{Bunn} and Y.~{Mao}, \emph{{Constraining f(R)
  gravity as a scalar-tensor theory}},
  \href{https://doi.org/10.1103/PhysRevD.76.063505}{\emph{\prd} {\bfseries 76}
  (2007) 063505} [\href{https://arxiv.org/abs/astro-ph/0612569}{{\ttfamily
  astro-ph/0612569}}].

\bibitem{Starobinsky2007f(R)ChameleonModels}
A.A.~{Starobinsky}, \emph{{Disappearing cosmological constant in f( R)
  gravity}}, \href{https://doi.org/10.1134/S0021364007150027}{\emph{Soviet
  Journal of Experimental and Theoretical Physics Letters} {\bfseries 86}
  (2007) 157} [\href{https://arxiv.org/abs/0706.2041}{{\ttfamily 0706.2041}}].

\bibitem{NavarroVanAcoleyen2007f(R)ChameleonModels}
I.~{Navarro} and K.~{Van Acoleyen}, \emph{{f(R) actions, cosmic acceleration
  and local tests of gravity}},
  \href{https://doi.org/10.1088/1475-7516/2007/02/022}{\emph{\jcap} {\bfseries
  2007} (2007) 022} [\href{https://arxiv.org/abs/gr-qc/0611127}{{\ttfamily
  gr-qc/0611127}}].

\bibitem{HuSawicki2007}
W.~{Hu} and I.~{Sawicki}, \emph{{Models of f(R) cosmic acceleration that evade
  solar system tests}},
  \href{https://doi.org/10.1103/PhysRevD.76.064004}{\emph{\prd} {\bfseries 76}
  (2007) 064004} [\href{https://arxiv.org/abs/0705.1158}{{\ttfamily
  0705.1158}}].

\bibitem{NFW1996}
J.F.~{Navarro}, C.S.~{Frenk} and S.D.M.~{White}, \emph{{The Structure of Cold
  Dark Matter Halos}}, \href{https://doi.org/10.1086/177173}{\emph{\apj}
  {\bfseries 462} (1996) 563}
  [\href{https://arxiv.org/abs/astro-ph/9508025}{{\ttfamily
  astro-ph/9508025}}].

\bibitem{BinneyTremaine2008}
J.~{Binney} and S.~{Tremaine}, \emph{{Galactic Dynamics: Second Edition}},
  Princeton University Press, Princeton, NJ USA (2008).

\bibitem{McMillan2017}
P.J.~{McMillan}, \emph{{The mass distribution and gravitational potential of
  the Milky Way}}, \href{https://doi.org/10.1093/mnras/stw2759}{\emph{\mnras}
  {\bfseries 465} (2017) 76}
  [\href{https://arxiv.org/abs/1608.00971}{{\ttfamily 1608.00971}}].

\bibitem{Deason2020}
A.J.~{Deason}, A.~{Fattahi}, C.S.~{Frenk}, R.J.J.~{Grand}, K.A.~{Oman},
  S.~{Garrison-Kimmel} et~al., \emph{{The edge of the Galaxy}},
  \href{https://doi.org/10.1093/mnras/staa1711}{\emph{\mnras} {\bfseries 496}
  (2020) 3929} [\href{https://arxiv.org/abs/2002.09497}{{\ttfamily
  2002.09497}}].

\bibitem{Adelberger:2009zz}
E.G.~Adelberger, J.H.~Gundlach, B.R.~Heckel, S.~Hoedl and S.~Schlamminger,
  \emph{{Torsion balance experiments: A low-energy frontier of particle
  physics}}, \href{https://doi.org/10.1016/j.ppnp.2008.08.002}{\emph{Prog.
  Part. Nucl. Phys.} {\bfseries 62} (2009) 102}.

\bibitem{Callingham+2019MWMass}
T.M.~{Callingham}, M.~{Cautun}, A.J.~{Deason}, C.S.~{Frenk}, W.~{Wang},
  F.A.~{G{\'o}mez} et~al., \emph{{The mass of the Milky Way from satellite
  dynamics}}, \href{https://doi.org/10.1093/mnras/stz365}{\emph{\mnras}
  {\bfseries 484} (2019) 5453}
  [\href{https://arxiv.org/abs/1808.10456}{{\ttfamily 1808.10456}}].

\bibitem{KesdenKamionkowski2006Streams1}
M.~{Kesden} and M.~{Kamionkowski}, \emph{{Galilean Equivalence for Galactic
  Dark Matter}},
  \href{https://doi.org/10.1103/PhysRevLett.97.131303}{\emph{\prl} {\bfseries
  97} (2006) 131303} [\href{https://arxiv.org/abs/astro-ph/0606566}{{\ttfamily
  astro-ph/0606566}}].

\bibitem{KesdenKamionkowski2006Streams2}
M.~{Kesden} and M.~{Kamionkowski}, \emph{{Tidal tails test the equivalence
  principle in the dark-matter sector}},
  \href{https://doi.org/10.1103/PhysRevD.74.083007}{\emph{\prd} {\bfseries 74}
  (2006) 083007} [\href{https://arxiv.org/abs/astro-ph/0608095}{{\ttfamily
  astro-ph/0608095}}].

\bibitem{Keselman+2009Streams}
J.A.~{Keselman}, A.~{Nusser} and P.J.E.~{Peebles}, \emph{{Galaxy satellites and
  the weak equivalence principle}},
  \href{https://doi.org/10.1103/PhysRevD.80.063517}{\emph{\prd} {\bfseries 80}
  (2009) 063517} [\href{https://arxiv.org/abs/0902.3452}{{\ttfamily
  0902.3452}}].

\bibitem{DuttonMaccio2014}
A.A.~{Dutton} and A.V.~{Macci{\`o}}, \emph{{Cold dark matter haloes in the
  Planck era: evolution of structural parameters for Einasto and NFW
  profiles}}, \href{https://doi.org/10.1093/mnras/stu742}{\emph{\mnras}
  {\bfseries 441} (2014) 3359}
  [\href{https://arxiv.org/abs/1402.7073}{{\ttfamily 1402.7073}}].

\bibitem{Moster2013}
B.P.~{Moster}, T.~{Naab} and S.D.M.~{White}, \emph{{Galactic star formation and
  accretion histories from matching galaxies to dark matter haloes}},
  \href{https://doi.org/10.1093/mnras/sts261}{\emph{\mnras} {\bfseries 428}
  (2013) 3121} [\href{https://arxiv.org/abs/1205.5807}{{\ttfamily 1205.5807}}].

\bibitem{Shen2003}
S.~{Shen}, H.J.~{Mo}, S.D.M.~{White}, M.R.~{Blanton}, G.~{Kauffmann},
  W.~{Voges} et~al., \emph{{The size distribution of galaxies in the Sloan
  Digital Sky Survey}},
  \href{https://doi.org/10.1046/j.1365-8711.2003.06740.x}{\emph{\mnras}
  {\bfseries 343} (2003) 978}
  [\href{https://arxiv.org/abs/astro-ph/0301527}{{\ttfamily
  astro-ph/0301527}}].

\bibitem{Bershady2010}
M.A.~{Bershady}, M.A.W.~{Verheijen}, K.B.~{Westfall}, D.R.~{Andersen},
  R.A.~{Swaters} and T.~{Martinsson}, \emph{{The DiskMass Survey. II. Error
  Budget}}, \href{https://doi.org/10.1088/0004-637X/716/1/234}{\emph{\apj}
  {\bfseries 716} (2010) 234}
  [\href{https://arxiv.org/abs/1004.5043}{{\ttfamily 1004.5043}}].

\end{thebibliography}\endgroup

\end{document}